\documentclass[10pt,a4paper,twocolumn]{article}
\usepackage[margin=1.72cm,columnsep=0.68cm]{geometry}
\usepackage{amsmath,amssymb,bm,graphicx,booktabs}
\usepackage[T1]{fontenc}
\usepackage{microtype}
\usepackage{dsfont}
\usepackage[sort,compress]{cite}
\usepackage{url}
\usepackage{xcolor}
\usepackage[hidelinks]{hyperref}
\usepackage{orcidlink}
\newcommand{\rev}[1]{#1}
\usepackage{balance}
\newcommand{\dd}{\mathrm{d}}
\newcommand{\ii}{\mathrm{i}}
\newcommand{\Lag}{\mathcal{L}}

\newcommand{\lKR}{\lambda_{\rm KR}}

\newenvironment{ruledtabular}{\centering}{\par}
\allowdisplaybreaks[1]
\begin{document}
\raggedbottom % Avoid stretching paragraph spacing to fill columns.
\title{Geometric Criticality and Quasibound Spectra of Massive Vector Fields in Kalb--Ramond Black Holes}
\author{\begin{tabular}{c}
Antonio Sousa Ribeiro \,\rev{\orcidlink{0000-0002-9570-6975}}\\
\small Instituto Federal do Piau\'i (IFPI), Campus S\~ao Raimundo Nonato, Brazil\\
\small \texttt{antonius.ribeirus@ifpi.edu.br}\\[4pt]
Francisco Artur Pinheiro Alves Filho \,\rev{\orcidlink{0000-0002-9723-9717}}\\
\small Universidade Federal do Vale do S\~ao Francisco (UNIVASF), Serra da Capivara, Brazil\\
\small \texttt{artur@univasf.edu.br}\\[4pt]
\rev{Lu\'is Rodolfo dos Santos Filho \,\orcidlink{0009-0001-5938-1165}}\\
\rev{\small Universidade Estadual Paulista (UNESP), Campus Guaratinguet\'a, Departamento de Matem\'atica, Brazil}\\
\rev{\small SEDUC -- Secretaria da Educa\c{c}\~ao do Estado de S\~ao Paulo, Brazil}
\end{tabular}}
\date{}
\maketitle
\begin{abstract}
\rev{We study Maxwell and Proca test fields in a static Lorentz-violating Kalb--Ramond black-hole geometry. The axial potential has an exact timelike-geodesic correspondence and a barrier--well fold at $r_{\rm fold}=6M(1-\lambda_{\rm KR})$ and $(M\mu)^2_{\rm c}=\ell(\ell+1)/[12(1-\lambda_{\rm KR})^2]$. Both parity systems admit exact Schwarzschild-form reductions with a continuously deformed angular parameter and rescaled spectral variables. For eight selected axial cases at $\ell=1,2,3$, real-energy barrier crossings precede their folds, at $0.8929\leq\mu_\times/\mu_c\leq0.9427$. Continued fractions, bilateral slope tests and independent Riccati shooting at all eight crossing couplings support local spectral regularity. Matrix-Riccati and QR--Evans matching track both electric-polarization families on 25 nodes of $0.38\le M\mu\le0.42$, $0\le\lambda_{\rm KR}\le0.10$. The smallest observed interfamily separation is $0.03305048/M$; this is not a continuous-domain bound, and no exceptional point is confirmed. Weak-binding continuation associates the selected axial branches with radial labels $n=0,1,0$ for $\ell=1,2,3$, respectively. The results isolate deformation-dependent spectral effects without claiming global completeness or a universal Lorentz-violation signature.}

\end{abstract}
\section{Introduction}
\label{sec:consolidated-1}

Massive Proca fields on black-hole geometries possess a longitudinal polarization and complex spectra determined by boundary conditions. The quasibound problem imposes ingoing waves at the horizon and exponential decay at infinity; it must be distinguished from the outgoing quasinormal and real-frequency scattering problems. An extremum merger of the real radial potential is not, by itself, a multiple zero of the complex spectral condition.

\rev{The Schwarzschild axial and polar Proca equations and spectra are established~\cite{Konoplya2006,RosaDolan2012,Ge2025}; the static Kalb--Ramond geometry, its geodesics and other perturbation sectors are also known~\cite{Yang2023,Junior2024,Periodic2025,Deng2025}. Related charged, dyonic, Dirac and rotating-scalar studies~\cite{GuKR2025,AraujoDyonic2026,Murodov2026,AydemirSakalli2026} involve different fields or backgrounds. Hod's inverted-potential analysis~\cite{Hod2025Inverted} and massive-scalar peak-resonance WKB treatment~\cite{Hod2026Scalar} pose distinct spectral problems. The latter motivates a potential-level comparison but does not predict low-multipole decaying Proca poles. Our earlier Schwarzschild study~\cite{RibeiroSchwarzschild} distinguished the axial fold from spectral degeneracy at zero deformation. Here the additional results are exact maps of both parity systems, finite-deformation axial benchmarks and energy-crossing diagnostics, and tracking of the coupled electric branches on a specified two-parameter grid. The fold--ISCO correspondence and generic fold exponents are not claimed as new.}

\rev{At fixed $M$ and $\ell$, the lapse $f=a-2M/r$, $a=(1-\lambda_{\rm KR})^{-1}$, changes the horizon, mass threshold and critical coupling. Frequencies are compared using the asymptotic clock $T=\sqrt a\,t$. In mapped variables, the change $\Lambda=\ell(\ell+1)/a$ cannot be absorbed into a frequency rescaling alone. At fixed normalized lapse mass $\mathcal M=M/a^{3/2}$ and field mass, this angular parameter is the only deformation in the mapped radial equations. All calculations concern minimally coupled test fields on the prescribed background, not dynamical perturbations of the Kalb--Ramond field.}

The paper is organized as follows. Section~\ref{sec:consolidated-2} presents the Kalb--Ramond geometry, Maxwell--Proca equations, and axial fold. Section~\ref{sec:consolidated-3} develops the axial quasibound spectrum and its numerical validation. Section~\ref{sec:consolidated-4} derives the coupled polar equations, while Sec.~\ref{sec:consolidated-5} studies the two electric-polarization families using matrix-Riccati and QR--Evans methods. Section~\ref{sec:consolidated-7} summarizes the conclusions. Appendix~\ref{app:matrix} records the matrix recurrence.

\section{Kalb--Ramond geometry, field equations, and axial criticality}
\label{sec:consolidated-2}
\subsection{Kalb--Ramond black-hole geometry}

\subsubsection{Metric and Lorentz-violation parameter}

We work in units
\begin{eqnarray}
 G&=&c=\hbar=1. \label{eq.2}
\end{eqnarray}
The static line element is
\begin{eqnarray}
 \dd s^2&=&-f(r)\dd t^2+\frac{\dd r^2}{f(r)}+r^2\left(\dd\theta^2+\sin^2\theta\,\dd\phi^2\right), \label{eq.3}
\end{eqnarray}
with
\begin{eqnarray}
 f(r)&=&a-\frac{2M}{r},\quad a\equiv\frac{1}{1-\lKR}. \label{eq.4}
\end{eqnarray}
The Schwarzschild limit is recovered when
\begin{eqnarray}
 \lKR&\longrightarrow&0,\quad a\longrightarrow1. \nonumber
\end{eqnarray}

Solving $f(r_{\rm h})=0$ gives
\begin{eqnarray}
 r_{\rm h}&=&\frac{2M}{a}=2M(1-\lKR).\label{eq.5}
\end{eqnarray}
Thus positive $\lambda_{\rm KR}$ moves the horizon inward in units of $M$, whereas negative $\lambda_{\rm KR}$ moves it outward.

\rev{The dimensionless parameter $\lambda_{\rm KR}$ encodes spontaneous Lorentz symmetry breaking in the gravitational sector: a nonzero vacuum expectation value of the antisymmetric Kalb--Ramond tensor, nonminimally coupled to curvature, selects a preferred local tensor configuration~\cite{Yang2023}. Its effect is not merely a mass or time rescaling. At fixed $M$, positive deformation decreases the horizon area and shifts the photon-sphere and ISCO radii inward, while the nonzero exterior Ricci curvature and the modified asymptotic spatial geometry persist after time normalization. The resulting changes in the effective barriers and polar couplings modify vector-field trapping and decay. Here these effects are mediated by the metric, without a direct Proca--Kalb--Ramond interaction.}
\begin{figure}[t]
\centering
\includegraphics[width=0.96\columnwidth]{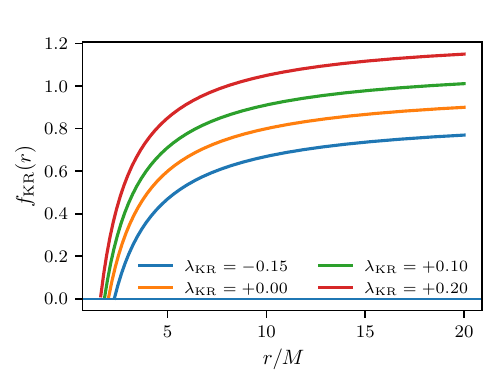}
\caption{Kalb--Ramond lapse function for representative Lorentz-violation parameters. The zero of each curve gives the event-horizon radius. Background lapse and horizon behavior have precedents in Kalb--Ramond black-hole geometry \cite{Yang2023,Junior2024}.}
\label{fig:lapse}
\end{figure}

\subsubsection{Asymptotic structure}

At large radius,
\begin{eqnarray}
 \lim_{r\rightarrow\infty}f(r)&=&a. \label{eq.6}
\end{eqnarray}
Hence
\begin{eqnarray}
 \dd s^2&\simeq&-a\,\dd t^2+a^{-1}\dd r^2+r^2\dd\Omega^2. \nonumber
\end{eqnarray}
One can normalize the temporal coefficient by introducing
\begin{eqnarray}
 T&=&\sqrt{a}\,t, \nonumber
\end{eqnarray}
but the radial and angular pieces cannot simultaneously be brought to standard Minkowski form while keeping $r$ as the area radius. This is the sense in which the geometry is not asymptotically Minkowskian in the standard spherical chart \cite{LiVlasov2024,ChargedKR2024}.

\rev{Introduce the normalized radial coordinate $\rho=r/\sqrt a$ and lapse mass scale $\mathcal M=M/a^{3/2}$. The exact metric becomes}
{\begin{eqnarray}
\dd s^2&=&-F_\rho\dd T^2+F_\rho^{-1}\dd\rho^2
+a\rho^2\dd\Omega^2,\nonumber\\
F_\rho&=&1-2\mathcal M/\rho.
\end{eqnarray}}
\rev{The asymptotic solid angle is $4\pi a$, an excess for positive deformation and a deficit for negative deformation. The scale $\mathcal M$ is defined by the normalized lapse, not assumed to be an ADM mass. At fixed $\mathcal M$ the horizon and fold in $\rho$ are $2\mathcal M$ and $6\mathcal M$, respectively, while their area radii retain the factor $\sqrt a$. Both parity maps have $\varpi=\mathcal M\Omega$, $\mathfrak m=\mathcal M\mu$ and $\Lambda=L/a$; at fixed normalized mass and field mass only the angular parameter changes in the mapped radial equations. In particular the normalized Hawking temperature is $1/(8\pi\mathcal M)$.}

\rev{Junior et al.~\cite{Junior2024} report a precession-based interval $-0.185022\leq\lambda_{\rm KR}\leq0.060938$ and a Gravity Probe B interval $[-6.30714,3.90708]\times10^{-12}$ under their adopted assumptions. Large deformations used here are theoretical sensitivity probes, not observationally allowed estimates; an observational comparison would require consistent mass calibration.}

The background is not Ricci flat; direct contraction of its curvature gives
\begin{eqnarray}
 R&=&-f''-\frac{4f'}{r}-\frac{2(f-1)}{r^2}
 =-\frac{2\lKR}{(1-\lKR)r^2}.\label{eq.7}
\end{eqnarray}
This observation matters because a derivation based solely on the Ricci-flat vector wave equation would be inappropriate. We shall instead derive the axial equation directly from the Proca field-strength equation.

\subsubsection{Tortoise coordinate}

Define the tortoise coordinate by
\begin{eqnarray}
 \frac{\dd r_*}{\dd r}&=&\frac{1}{f(r)}. \label{eq.8}
\end{eqnarray}
Integrating $\dd r_*/\dd r=1/f$ yields
\begin{eqnarray}
 r_*&=&\frac{r}{a}+\frac{2M}{a^2}\ln\left|\frac{ar}{2M}-1\right|.\label{eq.9}
\end{eqnarray}
As $r\rightarrow r_{\rm h}^{+}$, one has $r_*\rightarrow-\infty$.

\subsection{Maxwell and Proca fields}
\subsubsection{Field strength and Maxwell action}

Let $A_\mu$ be a vector potential and define
\begin{eqnarray}
 F_{\mu\nu}&=&\nabla_\mu A_\nu-\nabla_\nu A_\mu. \label{eq.10}
\end{eqnarray}
Because the connection is torsion free,
\begin{eqnarray}
 F_{\mu\nu}&=&\partial_\mu A_\nu-\partial_\nu A_\mu. \nonumber
\end{eqnarray}
The Maxwell action is
\begin{eqnarray}
 S_{\rm EM}&=&-\frac14\int \dd^4x\sqrt{-g}\,F_{\mu\nu}F^{\mu\nu}. \label{eq.11}
\end{eqnarray}
Its variation gives the Maxwell equations
\begin{eqnarray}
 \nabla_\mu F^{\mu\nu}&=&0.\label{eq.12}
\end{eqnarray}

\subsubsection{Proca action and dynamical Lorenz constraint}

The Proca action is
\begin{eqnarray}
 S_{\rm P}&=&\int\dd^4x\sqrt{-g}\left[-\frac14F_{\mu\nu}F^{\mu\nu}-\frac12\mu^2A_\mu A^\mu\right]. \label{eq.13}
\end{eqnarray}
Variation with respect to $A_\nu$ gives
\begin{eqnarray}
 \nabla_\mu F^{\mu\nu}-\mu^2A^\nu&=&0.\label{eq.14}
\end{eqnarray}
Taking the divergence and using the antisymmetry of $F^{\mu\nu}$ yields, for $\mu\ne0$,
\begin{eqnarray}
 \nabla_\mu A^\mu&=&0.\label{eq.15}
\end{eqnarray}
This is a dynamical constraint, not a gauge choice. Proca theory therefore propagates three physical polarizations rather than the two transverse polarizations of Maxwell theory.

One may combine Eqs.~\eqref{eq.14} and \eqref{eq.15} to write
\begin{eqnarray}
 \Box A^\nu-R^{\nu}{}_{\sigma}A^\sigma-\mu^2A^\nu&=&0. \label{eq.16}
\end{eqnarray}
Unlike Schwarzschild spacetime, the Kalb--Ramond geometry is not Ricci flat, so the explicit Ricci coupling in Eq.~\eqref{eq.16} must not be dropped in a component analysis. The axial derivation below avoids this ambiguity by starting from Eq.~\eqref{eq.14} itself.

\subsection{Vector spherical harmonics and parity}

Let $D_A$ be the covariant derivative on the unit sphere and $\epsilon_{AB}$ its Levi-Civita tensor. The scalar harmonics satisfy
\begin{eqnarray}
 D^AD_A Y_{\ell m}&=&-\ell(\ell+1)Y_{\ell m}. \label{eq.17}
\end{eqnarray}
Define the polar and axial vector harmonics by
\begin{eqnarray}
 Y_A^{\ell m}&=&D_A Y_{\ell m},\nonumber\\
 X_A^{\ell m}&=&-\epsilon_A{}^{B}D_B Y_{\ell m}. \label{eq.18}
\end{eqnarray}
The axial harmonic is divergence free,
\begin{eqnarray}
 D^AX_A^{\ell m}&=&0, \label{eq.19}
\end{eqnarray}
and has parity $(-1)^{\ell+1}$.

A general vector perturbation may be decomposed as
\begin{eqnarray}
 A_a&=&a_a(t,r)Y_{\ell m},\quad a\in\{t,r\},\nonumber\\
 A_A&=&a_{\rm e}(t,r)Y_A^{\ell m}+a_{\rm o}(t,r)X_A^{\ell m}. \label{eq.20}
\end{eqnarray}
The odd sector contains only the amplitude $a_{\rm o}$ and therefore provides the cleanest channel for isolating the effect of the Lorentz-violating background.

\subsection{Odd-parity Maxwell and Proca master equations}
\subsubsection{Odd ansatz and nonzero field-strength components}

For an axial mode we set
\begin{eqnarray}
 A_t&=&0,\quad A_r=0,\quad A_A=a(t,r)X_A^{\ell m}. \label{eq.21}
\end{eqnarray}
The nonzero mixed components of $F_{\mu\nu}$ are
\begin{eqnarray}
 F_{tA}&=&\partial_t a\,X_A^{\ell m},\nonumber\\
 F_{rA}&=&\partial_r a\,X_A^{\ell m},\nonumber\\
 F_{AB}&=&a\left(D_AX_B^{\ell m}-D_BX_A^{\ell m}\right). \nonumber
\end{eqnarray}
Because
\begin{eqnarray}
 g^{tt}&=&-\frac1f,\quad g^{rr}=f,\quad g^{AB}=\frac{1}{r^2}\Omega^{AB}, \nonumber
\end{eqnarray}
we have
\begin{eqnarray}
 F^{tA}&=&-\frac{1}{fr^2}\partial_ta\,X^{A},\nonumber\\
 F^{rA}&=&\frac{f}{r^2}\partial_ra\,X^{A}. \nonumber
\end{eqnarray}
The metric determinant is
\begin{eqnarray}
 \sqrt{-g}&=&r^2\sin\theta. \nonumber
\end{eqnarray}

\subsubsection{Angular projection}

The angular component $\nu=A$ of the Proca equation is
\begin{eqnarray}
 \nabla_\mu F^{\mu A}-\mu^2A^A&=&0. \nonumber
\end{eqnarray}
Using the divergence formula for an antisymmetric tensor,
\begin{eqnarray}
 \nabla_\mu F^{\mu A}&=&\frac{1}{\sqrt{-g}}\partial_\mu\left(\sqrt{-g}F^{\mu A}\right). \nonumber
\end{eqnarray}
The time contribution is
\begin{eqnarray}
 \frac{1}{\sqrt{-g}}\partial_t(\sqrt{-g}F^{tA})
 &=&-\frac{1}{fr^2}\partial_t^2a\,X^A. \nonumber
\end{eqnarray}
The radial contribution is
\begin{eqnarray}
 \frac{1}{\sqrt{-g}}\partial_r(\sqrt{-g}F^{rA})
 &=&\frac{1}{r^2}\partial_r\left(f\partial_ra\right)X^A. \nonumber
\end{eqnarray}
For the purely angular contribution, the standard axial-harmonic identity gives
\begin{eqnarray}
 D_B\left(D^BX^A-D^AX^B\right)&=&-\ell(\ell+1)X^A. \nonumber
\end{eqnarray}
Therefore
\begin{eqnarray}
 \frac{1}{r^4}D_B\left[r^2a(D^BX^A-D^AX^B)\right]
 &=&-\frac{\ell(\ell+1)}{r^4}aX^A. \nonumber
\end{eqnarray}
Finally,
\begin{eqnarray}
 A^A&=&\frac{a}{r^2}X^A. \nonumber
\end{eqnarray}
Multiplying the projected field equation by $r^2$ gives
\begin{eqnarray}
 -\frac{1}{f}\frac{\partial^2a}{\partial t^2}
 +\frac{\partial}{\partial r}\left(f\frac{\partial a}{\partial r}\right)
 -\left[\frac{\ell(\ell+1)}{r^2}+\mu^2\right]a&=&0. \label{eq.22}
\end{eqnarray}
This derivation uses only the field-strength Proca equation and remains valid although the Kalb--Ramond spacetime is not Ricci flat.

\subsubsection{Tortoise form}

From Eq.~\eqref{eq.8},
\begin{eqnarray}
 \frac{\partial}{\partial r_*}&=&f\frac{\partial}{\partial r},\nonumber\\
 \frac{\partial}{\partial r}&=&\frac{1}{f}\frac{\partial}{\partial r_*}. \nonumber
\end{eqnarray}
Hence
\begin{eqnarray}
 f\frac{\partial a}{\partial r}&=&\frac{\partial a}{\partial r_*},\nonumber\\
 \frac{\partial}{\partial r}\left(f\frac{\partial a}{\partial r}\right)
 &=&\frac{1}{f}\frac{\partial^2a}{\partial r_*^2}. \nonumber
\end{eqnarray}
Multiplying Eq.~\eqref{eq.22} by $f$ therefore yields
\begin{eqnarray}
 -\frac{\partial^2a}{\partial t^2}+\frac{\partial^2a}{\partial r_*^2}
 -f(r)\left[\frac{\ell(\ell+1)}{r^2}+\mu^2\right]a&=&0. \nonumber
\end{eqnarray}
For harmonic time dependence
\begin{eqnarray}
 a(t,r)&=&e^{-\ii\omega t}\psi(r), \nonumber
\end{eqnarray}
we obtain
\begin{eqnarray}
 \boxed{\frac{\dd^2\psi}{\dd r_*^2}+\left[\omega^2-V_{\rm P}^{(-)}(r)\right]\psi=0}, \label{eq.23}
\end{eqnarray}
where
\begin{eqnarray}
 \boxed{V_{\rm P}^{(-)}(r)=f_{\rm KR}(r)\left[\frac{\ell(\ell+1)}{r^2}+\mu^2\right]}. \label{eq.24}
\end{eqnarray}
The Maxwell odd-parity equation is obtained simply by setting $\mu=0$:
\begin{eqnarray}
 V_{\rm EM}(r)&=&f_{\rm KR}(r)\frac{\ell(\ell+1)}{r^2}. \label{eq.25}
\end{eqnarray}

\begin{figure}[t]
\centering
\includegraphics[width=0.96\columnwidth]{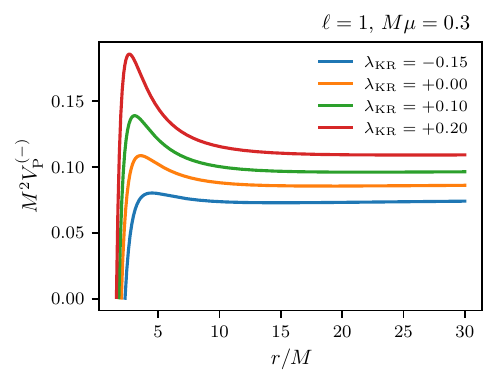}
\caption{Axial Proca potential at fixed multipole and vector mass for several values of the Kalb--Ramond Lorentz-violation parameter. The Schwarzschild massive-vector potential and Kalb--Ramond perturbative potentials have established antecedents; the displayed comparison concerns the specific mass and deformation parameters used here \cite{RosaDolan2012,Deng2025}.}
\label{fig:axpot}
\end{figure}

\subsection{Horizon and asymptotic behavior}

\subsubsection{Near the horizon}

As $r\rightarrow r_{\rm h}$,
\begin{eqnarray}
 f(r)&\longrightarrow&0,\nonumber\\
 V_{\rm P}^{(-)}&\longrightarrow&0. \nonumber
\end{eqnarray}
Equation~\eqref{eq.23} becomes
\begin{eqnarray}
 \frac{\dd^2\psi}{\dd r_*^2}+\omega^2\psi&\simeq&0. \nonumber
\end{eqnarray}
Therefore
\begin{eqnarray}
 \psi&\sim&e^{\pm\ii\omega r_*}. \nonumber
\end{eqnarray}
With the convention $e^{-\ii\omega t}$, the physical future-horizon solution is ingoing,
\begin{eqnarray}
 \boxed{\psi\sim e^{-\ii\omega r_*},\quad r_*\rightarrow-\infty}. \label{eq.26}
\end{eqnarray}

\subsubsection{Infinity and the shifted massive threshold}

At infinity,
\begin{eqnarray}
 f(r)&\longrightarrow&a,\quad \frac{\ell(\ell+1)}{r^2}\longrightarrow0. \nonumber
\end{eqnarray}
Hence
\begin{eqnarray}
 V_{\rm P}^{(-)}(r)&\longrightarrow&a\mu^2. \label{eq.27}
\end{eqnarray}
The asymptotic radial equation is
\begin{eqnarray}
 \frac{\dd^2\psi}{\dd r_*^2}+\left(\omega^2-a\mu^2\right)\psi&\simeq&0. \nonumber
\end{eqnarray}
Define
\begin{eqnarray}
 k_\infty^2&=&\omega^2-a\mu^2. \label{eq.28}
\end{eqnarray}
Oscillatory solutions require $k_\infty^2>0$. Exponentially decaying asymptotics require
\begin{eqnarray}
 \omega^2&<&a\mu^2. \label{eq.29}
\end{eqnarray}
Thus, in the original coordinate time, the massive threshold is
\begin{eqnarray}
 \boxed{\omega_{\rm th}=\sqrt{a}\,\mu=\frac{\mu}{\sqrt{1-\lKR}}}. \label{eq.30}
\end{eqnarray}
If one normalizes the asymptotic time by $T=\sqrt{a}\,t$, then the corresponding frequency is $\Omega=\omega/\sqrt{a}$ and the threshold becomes $\Omega=\mu$. This distinction is useful because it separates a coordinate normalization from the genuinely non-Minkowskian spatial asymptotics.

% Redundant figure omitted in consolidated presentation (available in original archive).

\subsection{Exact analytic structure of the axial potential}

\subsubsection{Dimensionless variables}

Introduce
\begin{eqnarray}
 x&=&\frac{r}{M},\quad \alpha=M\mu,\quad L=\ell(\ell+1). \label{eq.31}
\end{eqnarray}
Then
\begin{eqnarray}
 M^2V_{\rm P}^{(-)}(x)&=&\left(a-\frac{2}{x}\right)\left(\frac{L}{x^2}+\alpha^2\right). \label{eq.32}
\end{eqnarray}
We denote this dimensionless potential by $U(x)$.

\subsubsection{Stationary-point equation}

Differentiate Eq.~\eqref{eq.32}:
\begin{eqnarray}
 \frac{\dd U}{\dd x}
 &=&\frac{2}{x^2}\left(\frac{L}{x^2}+\alpha^2\right)
 +\left(a-\frac{2}{x}\right)\left(-\frac{2L}{x^3}\right). \nonumber
\end{eqnarray}
Expanding,
\begin{eqnarray}
 \frac{\dd U}{\dd x}
 &=&\frac{2L}{x^4}+\frac{2\alpha^2}{x^2}-\frac{2aL}{x^3}+\frac{4L}{x^4},\nonumber\\
 &=&\frac{2}{x^4}\left(\alpha^2x^2-aLx+3L\right). \nonumber
\end{eqnarray}
Thus stationary points satisfy
\begin{eqnarray}
 \boxed{\alpha^2x^2-aLx+3L=0}. \label{eq.33}
\end{eqnarray}
The two roots are
\begin{eqnarray}
 x_\pm&=&\frac{aL\pm\sqrt{a^2L^2-12L\alpha^2}}{2\alpha^2}. \label{eq.34}
\end{eqnarray}
Real extrema exist only if
\begin{eqnarray}
 a^2L^2-12L\alpha^2&\geq&0,\nonumber
\end{eqnarray}
or
\begin{eqnarray}
 \boxed{\alpha^2\leq\alpha_c^2\equiv\frac{a^2L}{12}}. \label{eq.35}
\end{eqnarray}
Equivalently,
\begin{eqnarray}
 \boxed{(M\mu)_c^2=\frac{\ell(\ell+1)}{12(1-\lKR)^2}}. \label{eq.36}
\end{eqnarray}
For $\alpha<\alpha_c$, the inner root $x_-$ is the barrier maximum continuously connected to the massless vector barrier, while $x_+$ is an outer minimum.

\begin{figure}[t]
\centering
\includegraphics[width=0.96\columnwidth]{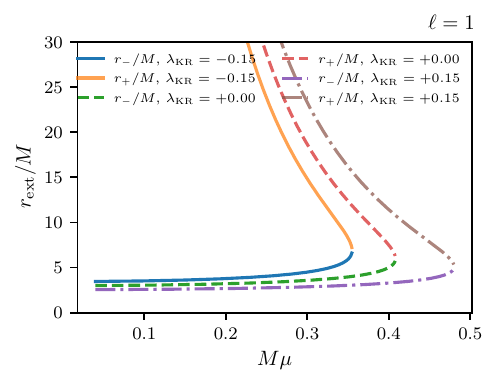}
\caption{Stationary-point branches of the axial Proca potential. Lorentz violation shifts both the fold coupling and the radial position of the merger.}
\label{fig:branches}
\end{figure}

\subsubsection{Fold radius}

At criticality the quadratic Eq.~\eqref{eq.33} has a double root. Using
\begin{eqnarray}
 \alpha_c^2&=&\frac{a^2L}{12}, \nonumber
\end{eqnarray}
we find
\begin{eqnarray}
 x_c&=&\frac{aL}{2\alpha_c^2},\nonumber\\
 &=&\frac{aL}{2(a^2L/12)},\nonumber\\
 &=&\frac{6}{a}. \nonumber
\end{eqnarray}
Therefore
\begin{eqnarray}
 \boxed{r_{\rm fold}=\frac{6M}{a}=6M(1-\lKR)}. \label{eq.37}
\end{eqnarray}
Since $r_{\rm h}=2M/a$,
\begin{eqnarray}
 \boxed{\frac{r_{\rm fold}}{r_{\rm h}}=3}. \label{eq.38}
\end{eqnarray}
The ratio is independent of the Lorentz-violation parameter.

\subsection{Timelike geodesics and exact potential correspondence}

\subsubsection{Geodesic effective potential}

Consider a unit-mass test particle in the equatorial plane $\theta=\pi/2$. Its Lagrangian is
\begin{eqnarray}
 2\Lag&=&-f\dot t^{\,2}+f^{-1}\dot r^{\,2}+r^2\dot\phi^{\,2}. \nonumber
\end{eqnarray}
Time translation and axial symmetry give conserved quantities
\begin{eqnarray}
 E&=&f\dot t,\quad \mathcal{J}=r^2\dot\phi. \label{eq.39}
\end{eqnarray}
For a timelike geodesic,
\begin{eqnarray}
 g_{\mu\nu}\dot x^\mu\dot x^\nu&=&-1. \nonumber
\end{eqnarray}
Substituting the metric,
\begin{eqnarray}
 -f\dot t^{\,2}+f^{-1}\dot r^{\,2}+r^2\dot\phi^{\,2}&=&-1. \nonumber
\end{eqnarray}
Using Eq.~\eqref{eq.39},
\begin{eqnarray}
 -f\left(\frac{E}{f}\right)^2+f^{-1}\dot r^{\,2}+r^2\left(\frac{\mathcal{J}}{r^2}\right)^2&=&-1,\nonumber\\
 -\frac{E^2}{f}+\frac{\dot r^{\,2}}{f}+\frac{\mathcal{J}^2}{r^2}&=&-1. \nonumber
\end{eqnarray}
Multiplying by $f$,
\begin{eqnarray}
 -E^2+\dot r^{\,2}+f\frac{\mathcal{J}^2}{r^2}&=&-f. \nonumber
\end{eqnarray}
Hence
\begin{eqnarray}
 \dot r^{\,2}+V_{\rm geo}(r)&=&E^2,\nonumber\\
 V_{\rm geo}(r)&=&f(r)\left(1+\frac{\mathcal{J}^2}{r^2}\right). \label{eq.40}
\end{eqnarray}

\subsubsection{Circular orbits and the ISCO}

A circular orbit satisfies
\begin{eqnarray}
 \frac{\dd V_{\rm geo}}{\dd r}&=&0. \nonumber
\end{eqnarray}
Differentiating,
\begin{eqnarray}
 \frac{\dd V_{\rm geo}}{\dd r}
 &=&f'\left(1+\frac{\mathcal{J}^2}{r^2}\right)-\frac{2f\mathcal{J}^2}{r^3}. \nonumber
\end{eqnarray}
Using $f'=2M/r^2$,
\begin{eqnarray}
 0&=&\frac{2M}{r^2}\left(1+\frac{\mathcal{J}^2}{r^2}\right)-\frac{2\mathcal{J}^2}{r^3}\left(a-\frac{2M}{r}\right). \nonumber
\end{eqnarray}
Multiplying by $r^4/2$,
\begin{eqnarray}
 0&=&Mr^2+M\mathcal{J}^2-a\mathcal{J}^2r+2M\mathcal{J}^2,\nonumber\\
 0&=&Mr^2-a\mathcal{J}^2r+3M\mathcal{J}^2. \label{eq.41}
\end{eqnarray}
The two circular-orbit branches merge when the discriminant vanishes:
\begin{eqnarray}
 a^2\mathcal{J}^4-12M^2\mathcal{J}^2&=&0. \nonumber
\end{eqnarray}
For nonzero angular momentum,
\begin{eqnarray}
 \mathcal{J}_{\rm ISCO}^2&=&\frac{12M^2}{a^2}. \nonumber
\end{eqnarray}
The corresponding radius is
\begin{eqnarray}
 \boxed{r_{\rm ISCO}=\frac{6M}{a}=6M(1-\lKR)}, \label{eq.42}
\end{eqnarray}
in agreement with the independent geodesic analyses of Refs.~\cite{Junior2024,Periodic2025}.

\subsubsection{Exact Proca--geodesic map}

Divide the Proca potential by $\mu^2$:
\begin{eqnarray}
 \frac{V_{\rm P}^{(-)}}{\mu^2}
 &=&f(r)\left[1+\frac{\ell(\ell+1)}{\mu^2r^2}\right]. \nonumber
\end{eqnarray}
Define
\begin{eqnarray}
 \mathcal{J}_{\rm eff}^2&=&\frac{\ell(\ell+1)}{\mu^2}. \label{eq.43}
\end{eqnarray}
Then
\begin{eqnarray}
 \boxed{\frac{V_{\rm P}^{(-)}}{\mu^2}=f(r)\left(1+\frac{\mathcal{J}_{\rm eff}^2}{r^2}\right)=V_{\rm geo}(r;\mathcal{J}_{\rm eff})}. \label{eq.44}
\end{eqnarray}
Thus the two functions of $r$ are exactly identical after the parameter identification Eq.~\eqref{eq.43}. This is a correspondence between effective potentials; it is not an identification of the wave equation with the geodesic equation.

Using Eq.~\eqref{eq.43} in the ISCO condition,
\begin{eqnarray}
 \frac{L}{\mu_c^2}&=&\frac{12M^2}{a^2},\nonumber
\end{eqnarray}
so
\begin{eqnarray}
 (M\mu_c)^2&=&\frac{a^2L}{12}, \nonumber
\end{eqnarray}
which is precisely Eq.~\eqref{eq.35}. Therefore
\begin{eqnarray}
 \boxed{r_{\rm fold}=r_{\rm ISCO}}. \label{eq.45}
\end{eqnarray}

\begin{figure}[t]
\centering
\includegraphics[width=0.96\columnwidth]{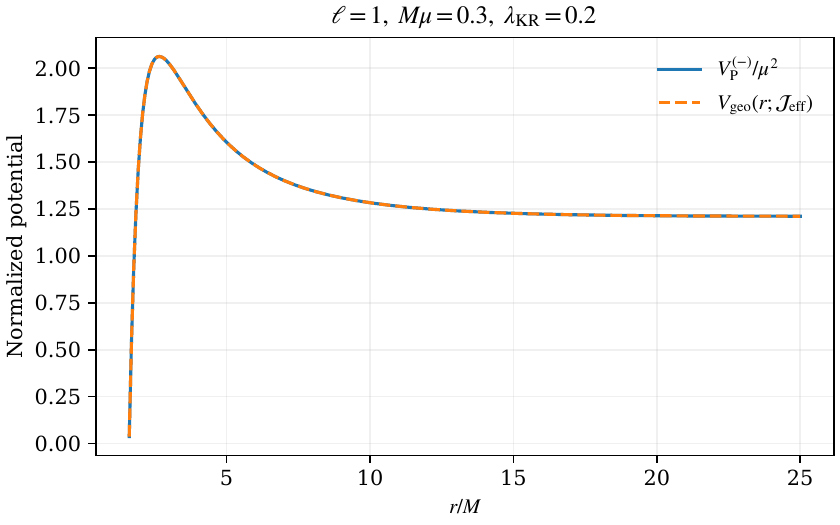}
\caption{\rev{Exact potential identity in Eq.~\eqref{eq.44} at $\ell=1$, $M\mu=0.3$ and $\lambda_{\rm KR}=0.2$. The two curves coincide.}}
\label{fig:geomap}
\end{figure}

The massless axial barrier also has a transparent geometric interpretation. For $\mu=0$,
\begin{eqnarray}
 V_{\rm EM}&=&f\frac{L}{r^2}. \nonumber
\end{eqnarray}
Its extremum satisfies
\begin{eqnarray}
 \frac{\dd}{\dd r}\left(\frac{f}{r^2}\right)&=&0. \nonumber
\end{eqnarray}
Explicitly,
\begin{eqnarray}
 \frac{f'}{r^2}-\frac{2f}{r^3}&=&0,\nonumber\\
 \frac{2M}{r^4}-\frac{2}{r^3}\left(a-\frac{2M}{r}\right)&=&0,\nonumber\\
 2M-2ar+4M&=&0,\nonumber\\
 \boxed{r_{\rm ph}=\frac{3M}{a}=3M(1-\lKR)}. \label{eq.46}
\end{eqnarray}
Thus the Maxwell-connected Proca maximum approaches the Kalb--Ramond photon sphere as $\mu\rightarrow0$.

\begin{figure}[t]
\centering
\includegraphics[width=0.96\columnwidth]{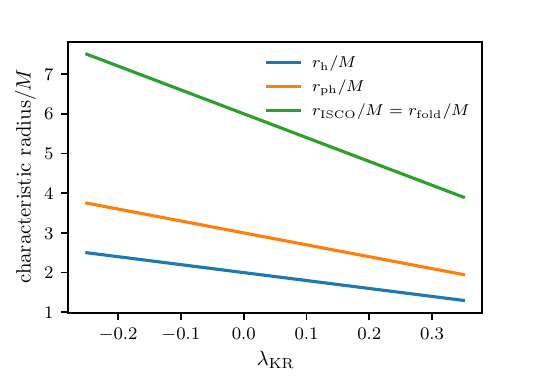}
\caption{Event horizon, photon sphere, and ISCO/fold radius. All three scale with the same factor $1-\lambda_{\rm KR}$. The characteristic orbital radii have literature precedents; the wave-potential fold identification is the present comparison \cite{Junior2024}.}
\label{fig:radii}
\end{figure}

\subsection{Exact critical scaling of the barrier--well fold}

\subsubsection{Critical parameter}

Define the reduced distance from the fold by
\begin{eqnarray}
 \delta&=&1-\frac{\alpha^2}{\alpha_c^2}
 =1-\frac{12\alpha^2}{a^2L},\quad 0\leq\delta\leq1. \label{eq.47}
\end{eqnarray}
Let
\begin{eqnarray}
 s&=&\sqrt{\delta}. \nonumber
\end{eqnarray}
Then
\begin{eqnarray}
 \alpha^2&=&\frac{a^2L}{12}(1-s^2). \nonumber
\end{eqnarray}
Substituting into Eq.~\eqref{eq.34},
\begin{eqnarray}
 x_-&=&\frac{aL-aL s}{2[a^2L(1-s^2)/12]},\nonumber\\
 &=&\frac{6(1-s)}{a(1-s^2)},\nonumber\\
 &=&\frac{6}{a(1+s)},\nonumber\\
 x_+&=&\frac{6}{a(1-s)}. \label{eq.48}
\end{eqnarray}
Therefore the separation is
\begin{eqnarray}
 \Delta x&=&x_+-x_-,\nonumber\\
 &=&\frac{6}{a}\left(\frac{1}{1-s}-\frac{1}{1+s}\right),\nonumber\\
 &=&\frac{6}{a}\left(\frac{2s}{1-s^2}\right),\nonumber\\
 \boxed{\Delta x=\frac{12\sqrt{\delta}}{a(1-\delta)}}. \label{eq.49}
\end{eqnarray}
Near the fold,
\begin{eqnarray}
 \Delta r&=&M\Delta x\simeq\frac{12M}{a}\delta^{1/2}+O(\delta^{3/2}). \label{eq.50}
\end{eqnarray}
Thus the stationary-point separation has the universal fold exponent $1/2$.

\subsubsection{Exact barrier--well depth}

At a stationary point, Eq.~\eqref{eq.33} implies
\begin{eqnarray}
 \alpha^2x^2&=&aLx-3L. \nonumber
\end{eqnarray}
Hence
\begin{eqnarray}
 \frac{L}{x^2}+\alpha^2
 &=&\frac{L+aLx-3L}{x^2},\nonumber\\
 &=&\frac{L(ax-2)}{x^2}. \nonumber
\end{eqnarray}
Also
\begin{eqnarray}
 a-\frac2x&=&\frac{ax-2}{x}. \nonumber
\end{eqnarray}
Therefore the potential at either extremum can be written as
\begin{eqnarray}
 \boxed{U_{\rm ext}=L\frac{(ax_{\rm ext}-2)^2}{x_{\rm ext}^3}}. \label{eq.51}
\end{eqnarray}
Using Eq.~\eqref{eq.48} and simplifying gives
\begin{eqnarray}
 U_-&=&\frac{La^3}{54}(2-s)^2(1+s),\nonumber\\
 U_+&=&\frac{La^3}{54}(1-s)(2+s)^2. \nonumber
\end{eqnarray}
Subtracting,
\begin{eqnarray}
 U_--U_+&=&\frac{La^3}{54}(2-s)^2(1+s)\nonumber\\
 &&-\frac{La^3}{54}(1-s)(2+s)^2,\nonumber\\
 &=&\frac{La^3}{27}s^3. \nonumber
\end{eqnarray}
Since $s^3=\delta^{3/2}$,
\begin{eqnarray}
 \boxed{M^2(V_{\max}-V_{\min})=\frac{La^3}{27}\delta^{3/2}}. \label{eq.52}
\end{eqnarray}
The exponent $3/2$ is unchanged by Lorentz violation, while the amplitude is rescaled by $a^3=(1-\lambda_{\rm KR})^{-3}$.

\begin{figure}[t]
\centering
\includegraphics[width=0.96\columnwidth]{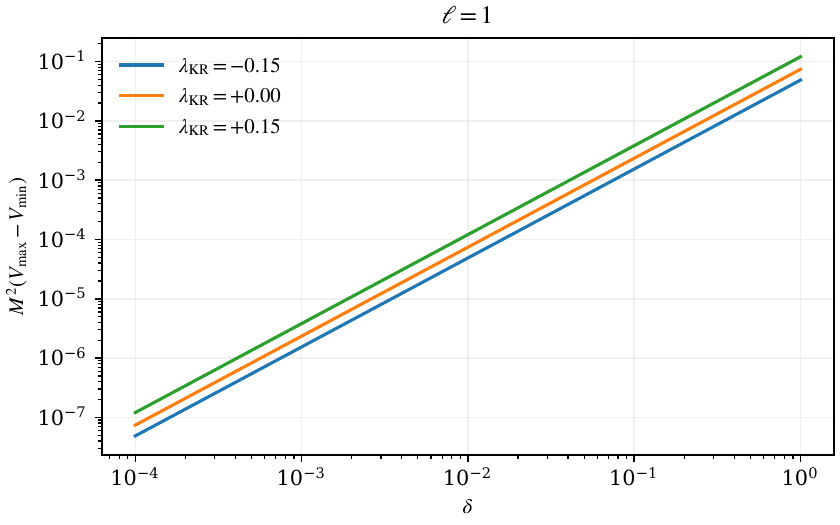}
\caption{Exact $\delta^{3/2}$ collapse of the axial barrier--well depth. The Lorentz-violating parameter changes the amplitude but not the critical exponent. The fold exponent is a generic local saddle-node scaling, rather than a claim of universality discovered here.}
\label{fig:depth}
\end{figure}

\subsubsection{Curvature of the extrema}

Let $x_*=r_*/M$. Since
\begin{eqnarray}
 \frac{\dd x_*}{\dd x}&=&\frac{1}{f},\quad \frac{\dd}{\dd x_*}=f\frac{\dd}{\dd x}, \nonumber
\end{eqnarray}
at an extremum, where $U_x=0$,
\begin{eqnarray}
 \frac{\dd^2U}{\dd x_*^2}
 &=&f\frac{\dd}{\dd x}\left(f\frac{\dd U}{\dd x}\right),\nonumber\\
 &=&f^2\frac{\dd^2U}{\dd x^2}+ff'\frac{\dd U}{\dd x},\nonumber\\
 &=&f^2U_{xx}. \nonumber
\end{eqnarray}
Substituting the exact branches yields
\begin{eqnarray}
 \left.\frac{\dd^2U}{\dd x_*^2}\right|_{x_-}
 &=&-\frac{La^7}{5832}s(2-s)^2(1+s)^4,\nonumber\\
 \left.\frac{\dd^2U}{\dd x_*^2}\right|_{x_+}
 &=&\frac{La^7}{5832}s(1-s)^4(2+s)^2. \label{eq.53}
\end{eqnarray}
For $s\ll1$,
\begin{eqnarray}
 \left.\frac{\dd^2U}{\dd x_*^2}\right|_{x_-}
 &\simeq&-\frac{La^7}{1458}\delta^{1/2}+O(\delta),\nonumber\\
 \left.\frac{\dd^2U}{\dd x_*^2}\right|_{x_+}
 &\simeq&+\frac{La^7}{1458}\delta^{1/2}+O(\delta). \label{eq.54}
\end{eqnarray}
Thus both extrema flatten with exponent $1/2$ as the fold is approached.

\section{Axial quasibound states and numerical validation}
\label{sec:consolidated-3}
\subsection{Boundary conditions and exact spectral reduction}

The axial equation can represent scattering, quasinormal modes, or quasibound states depending on the asymptotic conditions. Near the future event horizon, $r_*\rightarrow-\infty$, the physical mode is purely ingoing,
\begin{eqnarray}
 \psi&\sim&e^{-\ii\omega r_*}. \label{eq.56}
\end{eqnarray}
At infinity the potential approaches $a\mu^2$. For a quasibound state we define
\begin{eqnarray}
 \kappa&\equiv&\sqrt{a\mu^2-\omega^2},\quad \Re(\kappa)>0, \nonumber
\end{eqnarray}
and impose
\begin{eqnarray}
 \psi&\sim&e^{-\kappa r_*},\quad r_*\rightarrow+\infty. \label{eq.57}
\end{eqnarray}
For quasinormal modes the second condition is replaced by an outgoing wave with $k_\infty^2=\omega^2-a\mu^2$.

\subsubsection{Exact map to a Schwarzschild-form radial equation}

A useful simplification is obtained before introducing a Frobenius series. Define the dimensionless radial variable
\begin{eqnarray}
 R&\equiv&\frac{a r}{M}. \nonumber
\end{eqnarray}
Then
\begin{eqnarray}
 r&=&\frac{M}{a}R,\nonumber\\
 \frac{2M}{r}&=&\frac{2a}{R},\nonumber\\
 f(r)&=&a-\frac{2M}{r},\nonumber\\
 &=&a\left(1-\frac{2}{R}\right). \nonumber
\end{eqnarray}
We therefore introduce
\begin{eqnarray}
 F(R)&\equiv&1-\frac{2}{R},\quad f(r)=aF(R). \nonumber
\end{eqnarray}
Since
\begin{eqnarray}
 \frac{\dd r_*}{\dd r}&=&\frac{1}{f(r)},\nonumber\\
 \frac{\dd r}{\dd R}&=&\frac{M}{a},\nonumber
\end{eqnarray}
we have
\begin{eqnarray}
 \frac{\dd r_*}{\dd R}
 &=&\frac{\dd r_*}{\dd r}\frac{\dd r}{\dd R},\nonumber\\
 &=&\frac{1}{aF}\frac{M}{a},\nonumber\\
 &=&\frac{M}{a^2}\frac{1}{F}. \nonumber
\end{eqnarray}
Define a dimensionless tortoise coordinate $R_*$ by
\begin{eqnarray}
 R_*&\equiv&\frac{a^2}{M}r_*. \nonumber
\end{eqnarray}
Then
\begin{eqnarray}
 \frac{\dd R_*}{\dd R}&=&\frac{1}{F(R)},\nonumber\\
 R_*&=&R+2\ln\left(\frac{R}{2}-1\right)+C. \label{eq.58}
\end{eqnarray}
The derivative operators transform according to
\begin{eqnarray}
 \frac{\dd}{\dd r_*}&=&\frac{a^2}{M}\frac{\dd}{\dd R_*},\nonumber\\
 \frac{\dd^2}{\dd r_*^2}&=&\frac{a^4}{M^2}\frac{\dd^2}{\dd R_*^2}. \nonumber
\end{eqnarray}
Furthermore,
\begin{eqnarray}
 \frac{\ell(\ell+1)}{r^2}
 &=&\frac{a^2}{M^2}\frac{\ell(\ell+1)}{R^2}. \nonumber
\end{eqnarray}
Substituting these relations into
\begin{eqnarray}
 \frac{\dd^2\psi}{\dd r_*^2}+\left\{\omega^2-aF\left[\frac{a^2\ell(\ell+1)}{M^2R^2}+\mu^2\right]\right\}\psi&=&0,\nonumber
\end{eqnarray}
and multiplying by $M^2/a^4$, we obtain
\begin{eqnarray}
 \frac{\dd^2\psi}{\dd R_*^2}
 +\left[\varpi^2-F(R)\left(\mathfrak{m}^2+\frac{\Lambda}{R^2}\right)\right]\psi&=&0, \label{eq.59}
\end{eqnarray}
where the three effective spectral parameters are
\begin{eqnarray}
 \boxed{\varpi=\frac{M\omega}{a^2}},\quad
 \boxed{\mathfrak{m}=\frac{M\mu}{a^{3/2}}},\quad
 \boxed{\Lambda=\frac{\ell(\ell+1)}{a}}. \label{eq.60}
\end{eqnarray}
Equation~\eqref{eq.59} has exactly the algebraic form of the Schwarzschild odd-Proca equation with unit black-hole mass, except that the centrifugal eigenvalue is continuously deformed from $\ell(\ell+1)$ to $\Lambda$. The Schwarzschild problem is recovered term by term when $a\rightarrow1$.

The asymptotic bound-state momentum in these variables is
\begin{eqnarray}
 q&=&-\sqrt{\mathfrak{m}^2-\varpi^2},\quad \Re(-q)>0. \label{eq.61}
\end{eqnarray}
Indeed,
\begin{eqnarray}
 q^2&=&\mathfrak{m}^2-\varpi^2,\nonumber\\
 &=&\frac{M^2\mu^2}{a^3}-\frac{M^2\omega^2}{a^4},\nonumber\\
 &=&\frac{M^2}{a^4}\left(a\mu^2-\omega^2\right),\nonumber
\end{eqnarray}
so that
\begin{eqnarray}
 q&=&-\frac{M}{a^2}\kappa. \nonumber
\end{eqnarray}
Since $R=ar/M$, the exponential factor satisfies
\begin{eqnarray}
 e^{qR}&=&\exp\left(-\frac{\kappa}{a}r\right),\nonumber
\end{eqnarray}
which is precisely the leading large-$r$ behavior implied by $r_*\simeq r/a$.

\subsection{Frobenius expansion and continued fraction}

\subsubsection{Series variable and asymptotic factors}

Introduce the horizon-adapted variable
\begin{eqnarray}
 z&\equiv&F(R)=1-\frac{2}{R}. \label{eq.62}
\end{eqnarray}
The horizon is $z=0$ and spatial infinity is $z=1$. Solving for $R$ gives
\begin{eqnarray}
 R&=&\frac{2}{1-z}. \nonumber
\end{eqnarray}
The tortoise coordinate behaves near the horizon as
\begin{eqnarray}
 R_*&=&R+2\ln\left(\frac{R}{2}-1\right),\nonumber\\
 &\simeq&2\ln z+\hbox{constant},\quad z\rightarrow0. \nonumber
\end{eqnarray}
Thus the ingoing wave has the local behavior
\begin{eqnarray}
 e^{-\ii\varpi R_*}&\sim&z^{-2\ii\varpi}. \nonumber
\end{eqnarray}
At infinity the Coulomb-type power correction is written in terms of
\begin{eqnarray}
 \nu&\equiv&\frac{\varpi^2-q^2}{q}. \label{eq.63}
\end{eqnarray}
The full Frobenius ansatz is therefore
\begin{eqnarray}
 \boxed{\psi(R)=z^{-2\ii\varpi}R^{-\nu}e^{qR}\sum_{n=0}^{\infty}a_nz^n}. \label{eq.64}
\end{eqnarray}
For $a=1$, $R=r/M$ and Eq.~\eqref{eq.64} reduces exactly to the Schwarzschild ansatz used for the odd Proca quasibound spectrum.

\subsubsection{Derivation of the three-term recurrence}

To make the recurrence derivation explicit, first write Eq.~\eqref{eq.59} in the ordinary coordinate $R$. Since
\begin{eqnarray}
 \frac{\dd}{\dd R_*}&=&F\frac{\dd}{\dd R},\nonumber
\end{eqnarray}
we have
\begin{eqnarray}
 \frac{\dd^2\psi}{\dd R_*^2}
 &=&F\frac{\dd}{\dd R}\left(F\frac{\dd\psi}{\dd R}\right),\nonumber\\
 &=&F^2\frac{\dd^2\psi}{\dd R^2}+FF'\frac{\dd\psi}{\dd R}. \nonumber
\end{eqnarray}
Hence
\begin{eqnarray}
 F^2\psi_{,RR}+FF'\psi_{,R}
 +\left[\varpi^2-F\left(\mathfrak{m}^2+\frac{\Lambda}{R^2}\right)\right]\psi&=&0. \nonumber
\end{eqnarray}
Now
\begin{eqnarray}
 \frac{\dd z}{\dd R}&=&\frac{2}{R^2}=\frac{(1-z)^2}{2}. \nonumber
\end{eqnarray}
Therefore
\begin{eqnarray}
 \psi_{,R}&=&\frac{(1-z)^2}{2}\psi_{,z},\nonumber\\
 \psi_{,RR}&=&\frac{(1-z)^4}{4}\psi_{,zz}-\frac{(1-z)^3}{2}\psi_{,z}. \nonumber
\end{eqnarray}
Since $F=z$ and $F'=(1-z)^2/2$, the radial equation becomes
\begin{eqnarray}
 0&=&\frac{z^2(1-z)^4}{4}\psi_{,zz}\nonumber\\
 &&+\frac{z(1-z)^3(1-3z)}{4}\psi_{,z}\nonumber\\
 &&+\left[\varpi^2-z\mathfrak{m}^2-\frac{z\Lambda(1-z)^2}{4}\right]\psi. \nonumber
\end{eqnarray}
Write
\begin{eqnarray}
 \psi(z)&=&P(z)S(z),\nonumber\\
 P(z)&=&z^{-2\ii\varpi}(1-z)^{\nu}\exp\left(\frac{2q}{1-z}\right),\nonumber\\
 S(z)&=&\sum_{n=0}^{\infty}a_nz^n. \nonumber
\end{eqnarray}
The logarithmic derivative of $P$ is
\begin{eqnarray}
 H(z)&\equiv&\frac{P'}{P},\nonumber\\
 &=&-\frac{2\ii\varpi}{z}-\frac{\nu}{1-z}+\frac{2q}{(1-z)^2}. \nonumber
\end{eqnarray}
Thus
\begin{eqnarray}
 \psi'&=&P(S'+HS),\nonumber\\
 \psi''&=&P\left[S''+2HS'+(H'+H^2)S\right]. \nonumber
\end{eqnarray}
Substituting and using
\begin{eqnarray}
 q^2&=&\mathfrak{m}^2-\varpi^2,\quad \nu=\frac{\varpi^2-q^2}{q}, \nonumber
\end{eqnarray}
all fourth- and third-order poles at $z=1$ cancel. After division by the common factor $z(1-z)^2/4$, the equation for $S$ reduces to
\begin{eqnarray}
 0&=&z(1-z)^2S''+(c_0+c_1z+c_2z^2)S'\nonumber\\
 &&+(d_0+d_1z)S.\label{eq.65}
\end{eqnarray}
where
\begin{eqnarray}
 c_0&=&1-4\ii\varpi,\nonumber\\
 c_1&=&6q+8\ii\varpi-4-\frac{2\varpi^2}{q},\nonumber\\
 c_2&=&-2q-4\ii\varpi+3+\frac{2\varpi^2}{q},\nonumber\\
 d_0&=&-\Lambda-4q^2-12\ii q\varpi+3q+12\varpi^2+4\ii\varpi\nonumber\\
 &&+\frac{4\ii\varpi^3-\varpi^2}{q},\nonumber\\
 d_1&=&q^2+4\ii q\varpi-2q-6\varpi^2-4\ii\varpi\nonumber\\
 &&-\frac{4\ii\varpi^3}{q}+\frac{2\varpi^2}{q}+\frac{\varpi^4}{q^2}. \nonumber
\end{eqnarray}
Now
\begin{eqnarray}
 S'&=&\sum_{n=0}^{\infty}(n+1)a_{n+1}z^n,\nonumber\\
 S''&=&\sum_{n=0}^{\infty}(n+2)(n+1)a_{n+2}z^n. \nonumber
\end{eqnarray}
The coefficient of a generic power $z^n$ receives contributions proportional to $a_{n+1}$, $a_n$, and $a_{n-1}$. Collecting them gives
\begin{eqnarray}
 \alpha_0a_1+\beta_0a_0&=&0,\nonumber\\
 \alpha_na_{n+1}+\beta_na_n+\gamma_na_{n-1}&=&0,\quad n\geq1. \label{eq.66}
\end{eqnarray}
Directly from Eq.~\eqref{eq.65},
\begin{eqnarray}
 \alpha_n&=&(n+1)(n+c_0),\nonumber\\
 \beta_n&=&-2n(n-1)+c_1n+d_0,\nonumber\\
 \gamma_n&=&(n-1)(n-2)+c_2(n-1)+d_1. \nonumber
\end{eqnarray}
Substituting $c_i$ and $d_i$ and simplifying yields
{\footnotesize\begin{eqnarray}
 \alpha_n&=&(n+1)(n+1-4\ii\varpi), \label{eq.67}\\
 \beta_n&=&-2n^2+\frac{2}{q}\left(4\ii\varpi q+3q^2-\varpi^2-q\right)n\nonumber\\
 &&+\frac{1}{q}\left[3q^2+(12q-1)\varpi^2-4q^3-q\Lambda\right.\nonumber\\
 &&\left.\hspace{1.4cm}-12\ii\varpi q^2+4\ii\varpi q+4\ii\varpi^3\right], \nonumber\\
 \gamma_n&=&n^2-\frac{2}{q}\left(q^2-\varpi^2+2\ii\varpi q\right)n\nonumber\\
 &&-\frac{1}{q^2}\left[-4\ii\varpi q^3+4\ii\varpi^3q+6\varpi^2q^2-q^4-\varpi^4+q^2\right]. \label{eq.68}
\end{eqnarray}}
The only explicit imprint of the Kalb--Ramond deformation in the recurrence is through the effective triplet $(\varpi,\mathfrak{m},\Lambda)$ of Eq.~\eqref{eq.60}.

\subsubsection{Continued-fraction spectral equation}

For the minimal solution of the recurrence, the infinite continued fraction can be evaluated backward. At truncation order $N$ define
\begin{eqnarray}
 D_N&=&\beta_N,\nonumber\\
 D_n&=&\beta_n-\frac{\alpha_n\gamma_{n+1}}{D_{n+1}},\quad n=N-1,\ldots,1. \nonumber
\end{eqnarray}
The quasibound frequencies are roots of
\begin{eqnarray}
 \boxed{\mathcal{F}(\varpi,\mathfrak{m},\Lambda)\equiv
 \beta_0-\frac{\alpha_0\gamma_1}{D_1}=0}. \label{eq.69}
\end{eqnarray}
Equation~\eqref{eq.69}, together with Eq.~\eqref{eq.60}, is the desired continued-fraction spectral condition $\omega(M\mu,\lambda_{\rm KR})$.

\subsection{Independent axial benchmark with high-order shooting}\label{sec:v11-axial}

We use an independent scalar shooting test in addition to the perturbative and truncation audits. Let $Y=u_{r_*}/u$ be the axial logarithmic derivative. Differentiating and invoking the axial master equation gives
\begin{eqnarray}
\frac{\dd Y}{\dd r_*}&=&\frac{u_{r_*r_*}}{u}-\left(\frac{u_{r_*}}{u}\right)^2,\nonumber\\
&=&V_{\rm P}^{(-)}-\omega^2-Y^2.\label{eq:v11-riccati}
\end{eqnarray}
Since $\dd r_*/\dd R=F^{-1}$ in the Schwarzschild test with $R=r/M$, the chain rule implies
\begin{eqnarray}
\frac{\dd Y}{\dd R}&=&\frac{V_{\rm P}^{(-)}-\omega^2-Y^2}{F(R)}.\label{eq:v11-riccati-R}
\end{eqnarray}
Two solutions are propagated to $R_m=8$, one with the ingoing horizon condition and the other with the decaying asymptotic condition, and their difference $Y_{\rm H}(R_m)-Y_{\infty}(R_m)$ is required to vanish. \rev{The horizon series includes the constant angular contribution and is constructed by matching powers in Eq.~\eqref{eq:v11-riccati-R}, rather than fitting the eigenfunction.}

\begin{table}[ht]
\caption{Selected tests of the corrected independent axial shooting implementation for $\lambda_{\rm KR}=0$, $\ell=1$, and $M\mu=0.2$. The last column gives the observed complex-frequency difference from the scalar continued fraction at the specified numerical settings, not a rigorously certified error bound.}
\label{tab:v11-convergence}
\begin{ruledtabular}\scriptsize
\begin{tabular}{cccc}
Order & $R_{\max}$ & $\epsilon$ & $|\Delta(M\omega)|$\\
\hline
6 & 200 & $5\times10^{-4}$ & $2.46\times10^{-14}$\\
8 & 200 & $5\times10^{-4}$ & $2.36\times10^{-15}$\\
8 & 400 & $5\times10^{-4}$ & $5.38\times10^{-15}$\\
8 & 600 & $5\times10^{-4}$ & $4.97\times10^{-15}$\\
8 & 400 & $10^{-4}$ & $5.38\times10^{-15}$\\
8 & 400 & $10^{-3}$ & $5.38\times10^{-15}$
\end{tabular}
\end{ruledtabular}
\end{table}

With eighth-order asymptotic conditions, $R_{\max}=200$, $\epsilon=5\times10^{-4}$, and relative ODE tolerance $2\times10^{-10}$, the root is $M\omega_{\rm sh}\simeq0.198935455975584-1.77867167\times10^{-7}i$, whereas the continued fraction gives $M\omega_{\rm CF}\simeq0.198935455975584-1.77867169\times10^{-7}i$. The different boundary-value formulation and its stability under the displayed variations provide an independent numerical check of this \emph{single Schwarzschild axial benchmark}. The later sections supply selected finite-deformation axial checks and independent polar benchmarks; the full two-channel parameter space remains unvalidated.

\subsection{Quasibound spectrum and Schwarzschild comparison}

\subsubsection{Recovery of the Schwarzschild recurrence}

Setting $\lambda_{\rm KR}=0$ gives $a=1$. Therefore
\begin{eqnarray}
 \varpi&=&M\omega,\quad
 \mathfrak{m}=M\mu,\quad
 \Lambda=\ell(\ell+1). \nonumber
\end{eqnarray}
Equations~\eqref{eq.64}--\eqref{eq.69} then reduce identically to the Schwarzschild odd-Proca continued fraction. This provides an algebraic unit test independent of any root finding.

\subsubsection{Branch identity and spectral labeling}

\rev{Branch identity is established by backward continuation from the eight critical roots to $\mathfrak m=0.1\ell$, followed by forward continuation from the Coulomb estimate $\Re\Omega/\mu\simeq1-\mathfrak m^2/[2(n+\ell_{\rm eff}+1)^2]$, where $\ell_{\rm eff}=(\sqrt{1+4\Lambda}-1)/2$. This reduces to Rosa--Dolan's Eq.~(63) at zero deformation~\cite{RosaDolan2012}. With 301 nodes in each direction and truncations $N=900,1800$, the inferred radial labels are $n=0,1,0$ for $\ell=1,2,3$, respectively, at all sampled deformations. The effective radial index differs from these integers by at most $0.045$ at the weak-binding endpoints. The maximum forward/backward frequency discrepancy is $7.1\times10^{-10}$; the weakest damping rates are not resolved reliably in binary64. These associations use the real binding energy, not a global spectral count. In particular the selected quadrupole branch is not the fundamental. The leading damping law of Rosa--Dolan (their Eq.~64) applies at small coupling, not at the fold.}

\subsubsection{Critical frequencies}

We solve Eq.~\eqref{eq.69} with a backward truncation $N=450$ and continue roots continuously in $\lambda_{\rm KR}$. \rev{The crossing and slope checks below repeat the computation at $N=900$.} The physical coordinate frequency follows from
\begin{eqnarray}
 M\omega&=&a^2\varpi. \nonumber
\end{eqnarray}
Because $g_{tt}\rightarrow-a$ at infinity, it is also useful to introduce the asymptotically normalized time
\begin{eqnarray}
 T&=&\sqrt{a}\,t,\quad \Omega=\frac{\omega}{\sqrt{a}}. \label{eq.70}
\end{eqnarray}
In this normalization the massive threshold is simply $\Omega=\mu$.

\begin{figure}[t]
\centering
\includegraphics[width=0.96\columnwidth]{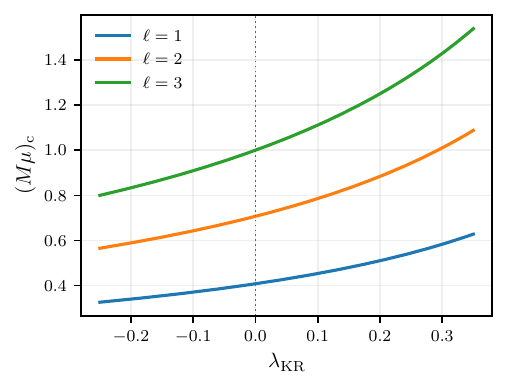}
\caption{Shift of the potential-fold coupling with the Kalb--Ramond parameter. The Schwarzschild values are recovered at $\lambda_{\rm KR}=0$.}
\label{fig:critcoupling-spectral}
\end{figure}

\begin{table*}[t]
\caption{Selected odd-parity quasibound roots at the shifted fold coupling; the weak-binding associations are $n=0,1,0$ for $\ell=1,2,3$. $\Omega=\omega/\sqrt{a}$ is the frequency referred to the asymptotically normalized time.}
\label{tab:critical-roots-kr}
\begin{ruledtabular}\scriptsize
\begin{tabular}{ccccccc}
$\ell$ & $\lambda_{\rm KR}$ & $(M\mu)_c$ & $\Re(M\omega_c)$ & $-\Im(M\omega_c)$ & $\Re(M\Omega_c)$ & $\Re(\Omega_c)/\mu$\\
\hline
1 & 0.0 & 0.408248 & 0.396145192 & $1.972908\times10^{-3}$ & 0.396145192 & 0.970354\\
1 & 0.1 & 0.453609 & 0.464509342 & $2.314467\times10^{-3}$ & 0.440672255 & 0.971480\\
1 & 0.2 & 0.510310 & 0.555001242 & $2.758027\times10^{-3}$ & 0.496408202 & 0.972757\\
1 & 0.3 & 0.583212 & 0.679103769 & $3.350152\times10^{-3}$ & 0.568178978 & 0.974224\\
2 & 0.0 & 0.707107 & 0.691726976 & $1.842593\times10^{-3}$ & 0.691726976 & 0.978250\\
2 & 0.2 & 0.883883 & 0.968780727 & $2.395230\times10^{-3}$ & 0.866503824 & 0.980337\\
3 & 0.0 & 1.000000 & 0.955129184 & $3.484938\times10^{-3}$ & 0.955129184 & 0.955129\\
3 & 0.2 & 1.250000 & 1.336842680 & $5.193579\times10^{-3}$ & 1.195708443 & 0.956567\\
\end{tabular}
\end{ruledtabular}
\end{table*}

\subsubsection{What changes physically with Lorentz violation?}

The Schwarzschild limit is $\lambda_{\rm KR}=0$ ($a=1$). At fixed $M$ and $\ell$, the geometric and wave-mechanical shifts obey
\begin{eqnarray}
\frac{r_{\rm h}(\lambda_{\rm KR})}{2M}&=&1-\lambda_{\rm KR},\nonumber\\
\frac{r_{\rm fold}(\lambda_{\rm KR})}{6M}&=&1-\lambda_{\rm KR},\nonumber\\
\frac{(M\mu)_c(\lambda_{\rm KR})}{(M\mu)_c(0)}&=&\frac{1}{1-\lambda_{\rm KR}},\nonumber\\
\frac{M^2V_c(\lambda_{\rm KR})}{M^2V_c(0)}&=&\frac{1}{(1-\lambda_{\rm KR})^3}.
\label{eq:kr-physical-shifts}
\end{eqnarray}
Here $V_c=2a^3\ell(\ell+1)/(27M^2)$ is the axial potential at its fold. For fixed dimensionless vector mass $m\equiv\alpha=M\mu$ and fixed area radius $x=r/M$ outside the respective horizons, the direct metric contribution satisfies
\begin{eqnarray}
\left.\frac{\partial(M^2V_-)}{\partial\lambda_{\rm KR}}\right|_{m,x}
&=&a^2\left(m^2+\frac{\ell(\ell+1)}{x^2}\right)>0.
\label{eq:kr-potential-shift}
\end{eqnarray}
This fixed-mass derivative should not be confused with comparing states at their respective critical couplings, for which $m_c$ itself changes. Moreover, the coordinate frequency and its asymptotically normalized counterpart obey $\Omega=\omega/\sqrt a$ and $\omega_{\rm th}=\sqrt a\,\mu$. These distinctions are necessary when comparing a Lorentz-violating black hole with Schwarzschild.

We also test whether the fold is an energy-level turning-point merger for the selected complex modes. With $w=M\omega$, define the real-energy diagnostic
\begin{eqnarray}
\mathcal B_c&=&\Re(w_c^2)-M^2V_c,\nonumber\\
&=&[\Re(w_c)]^2-[\Im(w_c)]^2
-\frac{2a^3\ell(\ell+1)}{27}.
\label{eq:kr-energy-gap}
\end{eqnarray}
\rev{All eight fold gaps in Table~\ref{tab:kr-energy-gap} are positive: the extrema disappear below the real modal energy proxy. The dimensionless ratio $\mathcal B_c/(M^2V_c)$ removes the overall $a^3$ scale and changes with deformation, even along the shifted critical curve. This comparison is not at fixed field mass. The real part of a squared complex frequency is a diagnostic, not a replacement for the complex boundary-value problem.}

\begin{table*}[t]
\centering\small
\caption{\rev{Axial fold gaps and real-energy barrier crossings. $d=-M\Im\Omega$ is the normalized-time damping. Crossings use the continued fraction at $N=900$; the fold gaps are from Table~\ref{tab:critical-roots-kr}.}}
\label{tab:kr-energy-gap}
\begin{tabular}{cccccc}
\hline
$\ell$ & $\lambda_{\rm KR}$ & $\mathcal B_c$ & $\mathcal B_c/(M^2V_c)$ & $m_\times/m_c$ & $d_c/d_\times$\\
\hline
1 & 0.0 & $8.77897\times10^{-3}$ & 0.05926 & 0.910452 & 2.683\\
1 & 0.1 & $1.25425\times10^{-2}$ & 0.06172 & 0.908473 & 2.654\\
1 & 0.2 & $1.86669\times10^{-2}$ & 0.06451 & 0.906274 & 2.621\\
1 & 0.3 & $2.92519\times10^{-2}$ & 0.06773 & 0.903804 & 2.583\\
2 & 0.0 & $3.40384\times10^{-2}$ & 0.07659 & 0.896061 & 4.921\\
2 & 0.2 & $7.04748\times10^{-2}$ & 0.08119 & 0.892887 & 4.636\\
3 & 0.0 & $2.33707\times10^{-2}$ & 0.02629 & 0.942689 & 3.133\\
3 & 0.2 & $5.10103\times10^{-2}$ & 0.02938 & 0.939013 & 3.082\\
\hline
\end{tabular}
\end{table*}

\rev{We locate earlier barrier crossings by continuing the selected roots at fixed deformation.} For $m<m_c$ the inner stationary point, with $L=\ell(\ell+1)$, is
\begin{eqnarray}
x_{\rm max}&=&\frac{6L}{aL+\sqrt{a^2L^2-12Lm^2}},
\label{eq:kr-barrier-max}
\end{eqnarray}
and a real-energy crossing satisfies
\begin{eqnarray}
\mathcal B(m,\lambda_{\rm KR})&=&\Re[(M\omega)^2]\nonumber\\
&&-M^2V_-(Mx_{\rm max})=0.
\label{eq:kr-barrier-cross}
\end{eqnarray}
\rev{Starting from the critical seeds, we continue to $m/m_c=0.84$ on 33 equally spaced points for each case, with $N=450,900$ and extrapolated frequency predictors. Each of the 16 tracks has one sampled sign change; a bracketed refinement re-solves the spectral condition at each trial mass. The largest sampled residual is $1.75\times10^{-11}$ and the crossing ratios agree within $3\times10^{-15}$ between truncations. All eight crossings precede the fold, and the normalized-time damping at the fold exceeds its crossing value (Table~\ref{tab:kr-energy-gap}). Independent Riccati shooting at each crossing mass, initialized with perturbed frequencies, recovers all eight poles at matching radii $R_m=8,10$ (outer radii 200, 300; series orders 8, 10). The maximum frequency difference is $5.46\times10^{-14}$ and the maximum matching residual is $6.36\times10^{-14}$; these checks validate the pole at the recorded coupling, not an independently located crossing. For each sampled $\ell$, increasing $\lambda_{\rm KR}$ lowers the crossing-to-fold mass ratio. This is a relative shift, not a fixed-$m$ comparison. We do not infer a sharp damping transition, global uniqueness of a crossing, or a polar turning-point criterion from this scalar diagnostic. These crossing calculations use the axial KR recurrence, not transferred Schwarzschild frequencies.}

\begin{figure*}[t]
\centering
\includegraphics[width=0.72\textwidth]{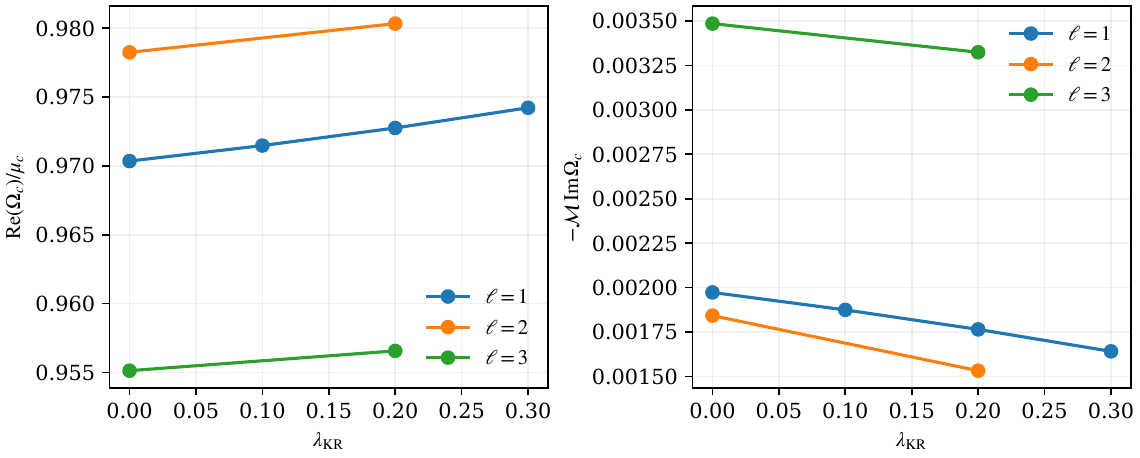}
\caption{\rev{Selected quasibound frequency at the shifted fold. Left: $\Re\Omega_c/\mu_c$. Right: $-\mathcal M\Im\Omega_c$. Points correspond to the eight tabulated cases; lines guide the eye.}}
\label{fig:critical-frequency-lambda}
\end{figure*}

\rev{Figure~\ref{fig:critical-frequency-lambda} removes both the clock and lapse-mass rescalings: it compares $\Re\Omega_c/\mu_c$ and $-\mathcal M\Im\Omega_c=-\Im\varpi_c$. Changes in these ratios reflect the mapped angular parameter at the respective folds, not a fixed-field-mass comparison.}

\begin{figure}[t]
\centering
\includegraphics[width=0.96\columnwidth]{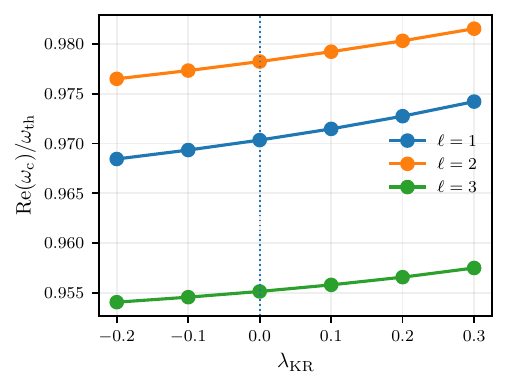}
\caption{Real part of the critical quasibound frequency relative to the shifted massive threshold. The roots remain below threshold throughout the representative deformation interval.}
\label{fig:threshold-ratio}
\end{figure}

The ratio to the shifted coordinate-time threshold
\begin{eqnarray}
 \omega_{\rm th}&=&\sqrt{a}\,\mu \nonumber
\end{eqnarray}
remains below unity for the branches displayed in Fig.~\ref{fig:threshold-ratio}. Thus the fold does not coincide with the asymptotic square-root branch point.

\subsubsection{Spectral response across the shifted fold}

Define, exactly as in the analytic potential problem,
\begin{eqnarray}
 \delta&=&1-\frac{(M\mu)^2}{(M\mu)_c^2}. \label{eq.71}
\end{eqnarray}
The potential stationary points obey $\Delta r\propto\delta^{1/2}$ and $V_{\max}-V_{\min}\propto\delta^{3/2}$. We test the complex frequency with
\begin{eqnarray}
 |\Re\omega-\Re\omega_c|&\propto&\delta^{p_R},\nonumber\\
 |\Im\omega-\Im\omega_c|&\propto&\delta^{p_I}. \label{eq.72}
\end{eqnarray}
For the selected $\ell=1$ branch the fitted exponents are
\begin{eqnarray}
 (p_R,p_I)_{\lambda_{\rm KR}=0}&=&(1.0017,0.9923),\nonumber\\
 (p_R,p_I)_{\lambda_{\rm KR}=0.1}&=&(1.0017,0.9925),\nonumber\\
 (p_R,p_I)_{\lambda_{\rm KR}=0.2}&=&(1.0017,0.9928),\nonumber\\
 (p_R,p_I)_{\lambda_{\rm KR}=0.3}&=&(1.0017,0.9931). \label{eq.73}
\end{eqnarray}
The Lorentz-violating deformation therefore leaves the local spectral exponent close to unity over the range studied.

\begin{figure*}[t]
\centering
\includegraphics[width=0.72\textwidth]{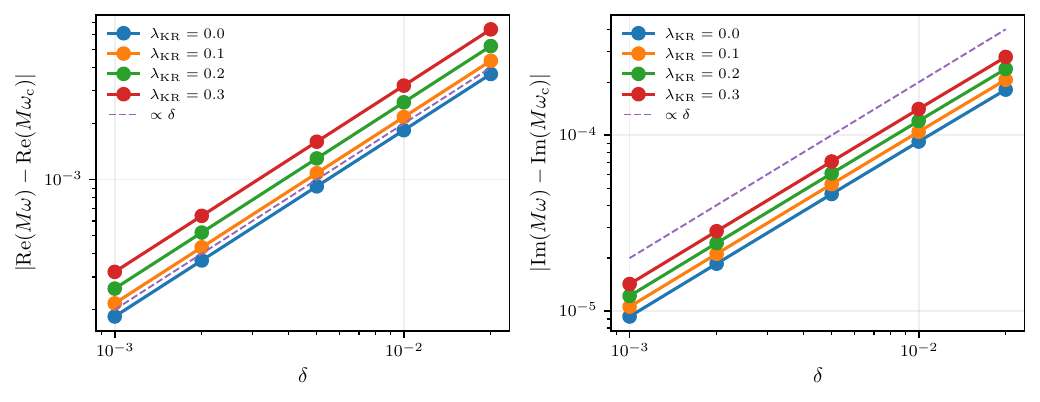}
\caption{Near-fold response of the selected $\ell=1$ quasibound root for Schwarzschild and representative Kalb--Ramond deformations. Both real and imaginary parts remain approximately linear in $\delta$.}
\label{fig:spectral-regularity}
\end{figure*}

\begin{figure}[t]
\centering
\includegraphics[width=0.69\columnwidth]{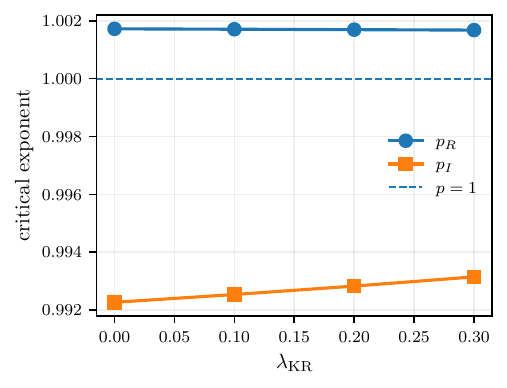}
\caption{Fitted critical exponents for the selected $\ell=1$ branch. The horizontal dashed line marks the analytic simple-root expectation $p=1$.}
\label{fig:spectral-exponents}
\end{figure}

The regularity follows from the same analytic mechanism as in Schwarzschild. At fixed $\ell$ and $\lambda_{\rm KR}$, the recurrence coefficients are analytic in $(\varpi,\mathfrak{m}^2)$ away from $q=0$. If the root is simple,
\begin{eqnarray}
 \left.\frac{\partial\mathcal{F}}{\partial\varpi}\right|_{c}&\neq&0, \label{eq.74}
\end{eqnarray}
then the implicit-function theorem gives a locally analytic branch. Numerically, for $\ell=1$ the magnitude $|\partial\mathcal{F}/\partial\varpi|$ at the fold changes smoothly from approximately $91.67$ at $\lambda_{\rm KR}=0$ to $93.62$ at $\lambda_{\rm KR}=0.3$, and is stable under variation of the finite-difference step. Thus the shifted fold remains a geometric nonanalyticity of the stationary-point parametrization, not a multiple spectral root.

\subsubsection{Two-sided derivative test at the axial fold}

\rev{To test local regularity, we differentiate the axial continued-fraction condition with respect to $s=(M\mu)^2$ at fixed $\ell$ and $\lambda_{\rm KR}$.} Writing $w=M\omega$ and $\mathcal F=\mathcal F(w,s;\lambda_{\rm KR})$, a simple root obeys
\begin{eqnarray}
 \frac{\dd w}{\dd s}&=&-
 \frac{\partial_s\mathcal F}{\partial_w\mathcal F},
 \quad \partial_w\mathcal F\ne0.\label{eq:fold-deriv-implicit}
\end{eqnarray}
As an independent finite-step estimator of the slope, roots were recomputed on both sides of $s_c=(M\mu_c)^2$:
\begin{eqnarray}
 D_h w&=&\frac{w(s_c+h)-w(s_c-h)}{2h},
 \quad h=\epsilon s_c.\label{eq:fold-deriv-central}
\end{eqnarray}
\rev{We extend the original centered test to all eight critical seeds, with $N=450,900$. At fixed deformation and $\ell$, the spectral partials are evaluated by centered differences with steps $10^{-6}$ in $w$ and $10^{-6}s_c$ in $s$. We independently recompute $w(s_c\pm h)$ and compare the implicit slope with both one-sided estimators, $[w(s_c+h)-w(s_c)]/h$ and $[w(s_c)-w(s_c-h)]/h$, as well as Eq.~\eqref{eq:fold-deriv-central}. Table~\ref{tab:two-sided-fold} summarizes 48 combinations of case, truncation and step.}

\begin{table*}[t]
\centering\small
\caption{\rev{Maximum absolute errors of the right, left and centered spectral-slope estimates relative to Eq.~\eqref{eq:fold-deriv-implicit}, over the eight cases and both truncations.}}
\label{tab:two-sided-fold}
\begin{tabular}{cccc}
\hline
$h/s_c$ & right & left & centered\\
\hline
$1.00\times10^{-3}$ & $3.22\times10^{-4}$ & $3.23\times10^{-4}$ & $2.10\times10^{-7}$\\
$1.00\times10^{-4}$ & $3.22\times10^{-5}$ & $3.22\times10^{-5}$ & $2.90\times10^{-8}$\\
$1.00\times10^{-5}$ & $3.22\times10^{-6}$ & $3.23\times10^{-6}$ & $2.97\times10^{-8}$\\
\hline
\end{tabular}
\end{table*}
\rev{The lateral errors decrease linearly with step; the centered error reaches a numerical floor from partial differentiation and root resolution. All evaluated $|\partial_w\mathcal F|$ exceed 45 in the chosen normalization, and the largest nearby-root residual is below $3\times10^{-12}$. The slope ratio is invariant under nonvanishing analytic rescalings at a root, whereas the derivative magnitude is not. These finite-fraction tests support local simple-root behavior but are not certified enclosures.}

\subsubsection{Independent Riccati-shooting validation}

The transformed equation also permits a direct shooting check without using the recurrence. Define the logarithmic derivative with respect to $R$,
\begin{eqnarray}
 y(R)&\equiv&\frac{1}{\psi}\frac{\dd\psi}{\dd R}. \nonumber
\end{eqnarray}
From
\begin{eqnarray}
 F^2\psi''+FF'\psi'+\left[\varpi^2-F\left(\mathfrak{m}^2+\frac{\Lambda}{R^2}\right)\right]\psi&=&0,\nonumber
\end{eqnarray}
and
\begin{eqnarray}
 \frac{\psi''}{\psi}&=&y'+y^2,\nonumber
\end{eqnarray}
we obtain
\begin{eqnarray}
 \boxed{\frac{\dd y}{\dd R}=-y^2-\frac{F'}{F}y-
 \frac{\varpi^2-F(\mathfrak{m}^2+\Lambda/R^2)}{F^2}}. \label{eq.75}
\end{eqnarray}
The horizon integration is initialized with the ingoing Frobenius branch and the outer integration with the decaying $e^{qR}$ branch. Matching the two logarithmic derivatives gives an independent complex residual. At the fold, selected modes agree with the continued-fraction roots to $8\times10^{-12}$ or better: for $(\ell,\lambda_{\rm KR})=(1,0.1)$ the difference in $M\omega$ is $5.5\times10^{-12}$, for $(1,0.2)$ it is $8.0\times10^{-12}$, and for $(2,0.2)$ it is $2.5\times10^{-13}$. This cross-check supports the continued-fraction implementation and the regular continuation through the shifted fold.

\section{Coupled even-parity Proca equations and spectral methods}
\label{sec:consolidated-4}
\subsection{Even-parity Proca sector: complete coupled derivation}

The odd-parity mode is special because a single axial vector harmonic carries the entire physical amplitude.  The even sector is qualitatively different.  A massive vector has a physical longitudinal polarization and, once gauge invariance is broken by the Proca mass, this polarization cannot be gauged away.  It is therefore useful to derive the coupled system directly from the field-strength equations rather than importing a Schwarzschild result by analogy.

\subsubsection{Even-parity harmonic ansatz}

Let
\begin{eqnarray}
 L&\equiv&\ell(\ell+1). \label{eq.83}
\end{eqnarray}
For a single Fourier-harmonic mode we take
\begin{eqnarray}
 A_t&=&\frac{u_1(r)}{r}Y_{\ell m}\,e^{-\ii\omega t},\nonumber\\
 A_r&=&\frac{u_2(r)}{r f(r)}Y_{\ell m}\,e^{-\ii\omega t},\nonumber\\
 A_A&=&\frac{u_3(r)}{L}D_A Y_{\ell m}\,e^{-\ii\omega t}. \label{eq.84}
\end{eqnarray}
The normalization in Eq.~\eqref{eq.84} is convenient because all three amplitudes have the same radial dimension and the angular eigenvalue appears only in simple algebraic combinations.

The relevant components of the field strength are obtained one by one.  First,
\begin{eqnarray}
 F_{tr}
 &=&\partial_tA_r-\partial_rA_t,\nonumber\\
 &=&-\ii\omega\frac{u_2}{rf}Y e^{-\ii\omega t}
 -\left(\frac{u_1'}{r}-\frac{u_1}{r^2}\right)Y e^{-\ii\omega t},\nonumber\\
 &=&\frac{1}{r}\left[-\frac{\ii\omega}{f}u_2-u_1'+\frac{u_1}{r}\right]Y e^{-\ii\omega t}. \label{eq.85}
\end{eqnarray}
Here and throughout this section a prime on $u_i$ denotes $\dd/\dd r$ unless a tortoise derivative is explicitly indicated.  It is useful to define
\begin{eqnarray}
 K(r)&\equiv&-\frac{\ii\omega}{f}u_2-u_1'+\frac{u_1}{r}. \label{eq.86}
\end{eqnarray}
Then
\begin{eqnarray}
 F_{tr}&=&\frac{K}{r}Y e^{-\ii\omega t}.\nonumber
\end{eqnarray}
The mixed temporal-angular component is
\begin{eqnarray}
 F_{tA}
 &=&\partial_tA_A-\partial_AA_t,\nonumber\\
 &=&\left(-\frac{\ii\omega}{L}u_3-\frac{u_1}{r}\right)D_A Y\,e^{-\ii\omega t}, \label{eq.87}
\end{eqnarray}
and the mixed radial-angular component is
\begin{eqnarray}
 F_{rA}
 &=&\partial_rA_A-\partial_AA_r,\nonumber\\
 &=&\left(\frac{u_3'}{L}-\frac{u_2}{rf}\right)D_A Y\,e^{-\ii\omega t}. \label{eq.88}
\end{eqnarray}
For the polar harmonic the purely angular antisymmetric component vanishes because $D_A D_B Y=D_B D_A Y$ for a scalar harmonic:
\begin{eqnarray}
 F_{AB}&=&0. \nonumber
\end{eqnarray}

\subsubsection{Lorenz constraint in radial variables}

The determinant of the metric is
\begin{eqnarray}
 \sqrt{-g}&=&r^2\sin\theta.\nonumber
\end{eqnarray}
The contravariant components of the potential are
\begin{eqnarray}
 A^t&=&-\frac{u_1}{rf}Y e^{-\ii\omega t},\nonumber\\
 A^r&=&\frac{u_2}{r}Y e^{-\ii\omega t},\nonumber\\
 A^A&=&\frac{u_3}{Lr^2}D^A Y e^{-\ii\omega t}.\nonumber
\end{eqnarray}
Using
\begin{eqnarray}
 \nabla_\mu A^\mu
 &=&\frac{1}{\sqrt{-g}}\partial_\mu\left(\sqrt{-g}A^\mu\right),\nonumber
\end{eqnarray}
we evaluate the three contributions separately:
\begin{eqnarray}
 \frac{1}{r^2}\partial_t(r^2A^t)
 &=&\frac{\ii\omega u_1}{rf}Y e^{-\ii\omega t},\nonumber\\
 \frac{1}{r^2}\partial_r(r^2A^r)
 &=&\frac{(ru_2)'}{r^2}Y e^{-\ii\omega t},\nonumber\\
 D_AA^A
 &=&\frac{u_3}{Lr^2}D_AD^AY e^{-\ii\omega t},\nonumber\\
 &=&-\frac{u_3}{r^2}Y e^{-\ii\omega t}.\nonumber
\end{eqnarray}
Therefore the dynamical Lorenz constraint becomes
\begin{eqnarray}
 \frac{\ii\omega u_1}{rf}+\frac{(ru_2)'}{r^2}-\frac{u_3}{r^2}&=&0,\nonumber\\
 \boxed{\frac{\ii\omega r}{f}u_1+(ru_2)'-u_3=0}. \label{eq.89}
\end{eqnarray}
Solving for the temporal amplitude,
\begin{eqnarray}
 u_1
 &=&\frac{f}{\ii\omega r}\left[u_3-(ru_2)'\right]. \label{eq.90}
\end{eqnarray}
Equation~\eqref{eq.90} makes explicit that $u_1$ is not an independent dynamical variable once the massive constraint has been imposed.

\subsubsection{The $t$ component of the Proca equation}

The temporal field equation, after harmonic projection, reads
\begin{eqnarray}
 (rK)'+\frac{1}{f}\left(\ii\omega u_3+\frac{Lu_1}{r}+\mu^2ru_1\right)&=&0. \label{eq.91}
\end{eqnarray}

\subsubsection{The $r$ component}

Using $F^{tr}=-K/r$ and
\begin{eqnarray}
 F^{Ar}
 &=&-\frac{f}{r^2}\left(\frac{u_3'}{L}-\frac{u_2}{rf}\right)D^A Y e^{-\ii\omega t},\nonumber
\end{eqnarray}
we obtain
\begin{eqnarray}
 \nabla_\mu F^{\mu r}
 &=&\partial_tF^{tr}+D_AF^{Ar},\nonumber\\
 &=&\frac{\ii\omega K}{r}Y e^{-\ii\omega t}
 +\left(\frac{fu_3'}{r^2}-\frac{Lu_2}{r^3}\right)Y e^{-\ii\omega t}.\nonumber
\end{eqnarray}
Moreover,
\begin{eqnarray}
 -\mu^2A^r&=&-\frac{\mu^2u_2}{r}Y e^{-\ii\omega t}.\nonumber
\end{eqnarray}
The radial Proca equation is therefore
\begin{eqnarray}
 \boxed{\ii\omega K+\frac{f}{r}u_3'-\left(\frac{L}{r^2}+\mu^2\right)u_2=0}. \label{eq.92}
\end{eqnarray}

\subsubsection{The angular equation}

For an angular index $A$ the nonzero contributions are the temporal and radial divergences.  From Eq.~\eqref{eq.87},
\begin{eqnarray}
 F^{tA}
 &=&\frac{1}{fr^2}\left(\frac{\ii\omega}{L}u_3+\frac{u_1}{r}\right)D^AY e^{-\ii\omega t}.\nonumber
\end{eqnarray}
Hence
\begin{eqnarray}
 \partial_tF^{tA}
 &=&\frac{1}{fr^2}\left(\frac{\omega^2}{L}u_3-\frac{\ii\omega u_1}{r}\right)D^AY e^{-\ii\omega t}.\nonumber
\end{eqnarray}
The radial term is
\begin{eqnarray}
 \frac{1}{r^2}\partial_r(r^2F^{rA})
 &=&\frac{1}{r^2}\frac{\dd}{\dd r}\left(\frac{fu_3'}{L}-\frac{u_2}{r}\right)D^AY e^{-\ii\omega t}.\nonumber
\end{eqnarray}
Finally,
\begin{eqnarray}
 -\mu^2A^A&=&-\frac{\mu^2u_3}{Lr^2}D^AY e^{-\ii\omega t}.\nonumber
\end{eqnarray}
Multiplying the angular equation by $Lr^2$ yields
\begin{eqnarray}
 0&=&\frac{\dd}{\dd r}(fu_3')
 -L\frac{\dd}{\dd r}\left(\frac{u_2}{r}\right)\nonumber\\
 &&+\left(\frac{\omega^2}{f}-\mu^2\right)u_3
 -\frac{\ii\omega L}{fr}u_1.\label{eq.93}
\end{eqnarray}
Equations~\eqref{eq.89}, \eqref{eq.91}, \eqref{eq.92}, and \eqref{eq.93} form a redundant but useful representation of the complete even-parity system.  Only two second-order dynamical equations are independent after the Lorenz constraint is used.

\subsubsection{Reduction to two coupled tortoise equations}

Introduce the tortoise derivative
\begin{eqnarray}
 \frac{\dd}{\dd r_*}&=&f\frac{\dd}{\dd r}.\nonumber
\end{eqnarray}
For harmonic time dependence define the common spin--1 radial operator
\begin{eqnarray}
 \mathcal D_1
 &=&\frac{\dd^2}{\dd r_*^2}+\omega^2
 -f\left(\frac{L}{r^2}+\mu^2\right). \label{eq.94}
\end{eqnarray}
Substitution of Eq.~\eqref{eq.90} into Eqs.~\eqref{eq.92} and \eqref{eq.93}, followed by repeated use of
\begin{eqnarray}
 \frac{\dd f}{\dd r_*}&=&ff',\nonumber
\end{eqnarray}
reduces the system to
\begin{eqnarray}
 \mathcal D_1u_2
 +\frac{f}{r}\left(f'-\frac{2f}{r}\right)(u_2-u_3)&=&0, \label{eq.95}\\
 \mathcal D_1u_3
 +\frac{2fL}{r^2}u_2&=&0. \label{eq.96}
\end{eqnarray}
For the present Kalb--Ramond lapse,
\begin{eqnarray}
 f'&=&\frac{2M}{r^2},\nonumber
\end{eqnarray}
so the first coupling coefficient becomes
\begin{eqnarray}
 \frac{f}{r}\left(f'-\frac{2f}{r}\right)
 &=&\frac{f}{r}\left[\frac{2M}{r^2}-\frac{2}{r}\left(a-\frac{2M}{r}\right)\right],\nonumber\\
 &=&\frac{2f}{r^3}(3M-ar). \label{eq.97}
\end{eqnarray}
The second coupling is simply
\begin{eqnarray}
 \frac{2fL}{r^2}. \nonumber
\end{eqnarray}
The system may therefore be written in matrix form as
\begin{eqnarray}
 \frac{\dd^2}{\dd r_*^2}
 \left(\begin{array}{c}u_2\\u_3\end{array}\right)
 +\left[\omega^2\mathds{1}_{2}-\mathbf V_{\rm even}(r)\right]
 \left(\begin{array}{c}u_2\\u_3\end{array}\right)&=&0, \label{eq.98}
\end{eqnarray}
where
\begin{eqnarray}
 \mathbf V_{\rm even}
 &=&\left(\begin{array}{cc}
 V_0-C & C\\
 -D & V_0
 \end{array}\right),\nonumber\\
 V_0&=&f\left(\frac{L}{r^2}+\mu^2\right),\nonumber\\
 C&=&\frac{2f}{r^3}(3M-ar),\nonumber\\
 D&=&\frac{2fL}{r^2}. \label{eq.99}
\end{eqnarray}
The matrix in Eq.~\eqref{eq.99} is not symmetric in the variables $(u_2,u_3)$ because these amplitudes are not canonically normalized orthogonal polarization coordinates.  No physical instability should be inferred from the pointwise eigenvalues of this coefficient matrix alone; the spectrum is a global boundary-value problem.

\begin{figure}[t]
\centering
\includegraphics[width=0.98\columnwidth]{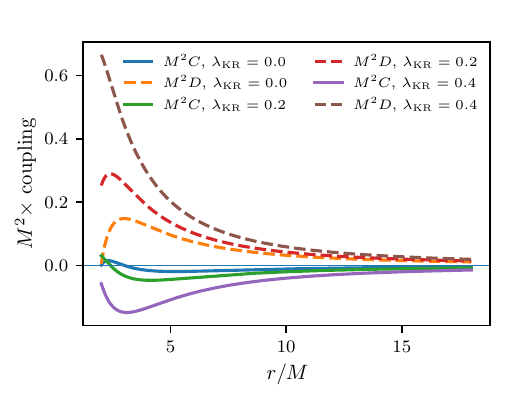}
\caption{Even-parity coupling functions in the original area radius for $\ell=1$.  Lorentz violation changes both the horizon position and the radial profile of the couplings.  Solid curves show $M^2C$ and dashed curves show $M^2D$. The polar-coupling structure is adapted from the Schwarzschild Proca formulation, with the Kalb--Ramond metric inserted \cite{RosaDolan2012}.}
\label{fig:evenphysical}
\end{figure}

\subsubsection{Exact reduction of the even system to Schwarzschild form}

The axial sector admitted an exact rescaling to a Schwarzschild-form equation.  Remarkably, the same transformation also simplifies the complete even system.  Define
\begin{eqnarray}
 R&=&\frac{ar}{M},\quad R_*\equiv\frac{a^2r_*}{M},\nonumber\\
 \varpi&=&\frac{M\omega}{a^2},\quad
 \mathfrak m\equiv\frac{M\mu}{a^{3/2}},\quad
 \Lambda\equiv\frac{L}{a}. \label{eq.100}
\end{eqnarray}
Then
\begin{eqnarray}
 f(r)&=&aF(R),\quad F(R)=1-\frac{2}{R}.\nonumber
\end{eqnarray}
The derivative transforms as
\begin{eqnarray}
 \frac{\dd}{\dd r_*}&=&\frac{a^2}{M}\frac{\dd}{\dd R_*},\nonumber\\
 \frac{\dd^2}{\dd r_*^2}&=&\frac{a^4}{M^2}\frac{\dd^2}{\dd R_*^2}.\nonumber
\end{eqnarray}
For the common potential term,
\begin{eqnarray}
 \frac{M^2}{a^4}f\left(\frac{L}{r^2}+\mu^2\right)
 &=&\frac{M^2}{a^4}aF\left(\frac{La^2}{M^2R^2}+\mu^2\right),\nonumber\\
 &=&F\left(\frac{L/a}{R^2}+\frac{M^2\mu^2}{a^3}\right),\nonumber\\
 &=&F\left(\frac{\Lambda}{R^2}+\mathfrak m^2\right).\nonumber
\end{eqnarray}
The first coupling transforms as
\begin{eqnarray}
 \frac{M^2}{a^4}C
 &=&\frac{M^2}{a^4}\frac{2f}{r^3}(3M-ar),\nonumber\\
 &=&\frac{M^2}{a^4}\frac{2aF}{(MR/a)^3}M(3-R),\nonumber\\
 &=&\frac{2F(3-R)}{R^3}.\nonumber
\end{eqnarray}
The second becomes
\begin{eqnarray}
 \frac{M^2}{a^4}D
 &=&\frac{M^2}{a^4}\frac{2aFL}{M^2R^2/a^2},\nonumber\\
 &=&\frac{2\Lambda F}{R^2}.\nonumber
\end{eqnarray}
Dividing Eqs.~\eqref{eq.95}--\eqref{eq.96} by $a^4/M^2$ gives
\begin{eqnarray}
 \widehat{\mathcal D}_1u_2
 +\frac{2F(3-R)}{R^3}(u_2-u_3)&=&0, \label{eq.101}\\
 \widehat{\mathcal D}_1u_3
 +\frac{2\Lambda F}{R^2}u_2&=&0, \label{eq.102}
\end{eqnarray}
with
\begin{eqnarray}
 \widehat{\mathcal D}_1
 &=&\frac{\dd^2}{\dd R_*^2}+\varpi^2
 -F\left(\frac{\Lambda}{R^2}+\mathfrak m^2\right). \label{eq.103}
\end{eqnarray}
Equations~\eqref{eq.101}--\eqref{eq.103} are algebraically identical to the Schwarzschild even-parity Proca system, except that the angular eigenvalue has been continuously deformed from the integer-spherical value $L$ to $\Lambda=L/a$.  This exact map is one of the central structural results of the extended analysis.

% Redundant figure omitted in consolidated presentation (available in original archive).

\subsubsection{Schwarzschild limit}

The limit $\lambda_{\rm KR}\rightarrow0$ implies
\begin{eqnarray}
 a&\rightarrow&1,\nonumber\\
 R&\rightarrow&r/M,\nonumber\\
 \varpi&\rightarrow&M\omega,\nonumber\\
 \mathfrak m&\rightarrow&M\mu,\nonumber\\
 \Lambda&\rightarrow&L.\nonumber
\end{eqnarray}
Therefore Eqs.~\eqref{eq.101}--\eqref{eq.102} reduce exactly to the coupled Schwarzschild equations used in the classic matrix-valued continued-fraction analysis of Rosa and Dolan.  This provides a nontrivial consistency check because it involves the derivative terms and both off-diagonal couplings, not merely the common diagonal potential.

\subsection{Near-horizon and asymptotic analysis of the even polarizations}

\subsubsection{Near-horizon indicial behavior}

In the scaled variables the horizon is always at
\begin{eqnarray}
 R_{\rm h}&=&2.\nonumber
\end{eqnarray}
Near the horizon set
\begin{eqnarray}
 R&=&2+\epsilon,\quad 0<\epsilon\ll1.\nonumber
\end{eqnarray}
Then
\begin{eqnarray}
 F(R)
 &=&1-\frac{2}{2+\epsilon},\nonumber\\
 &=&1-\frac{1}{1+\epsilon/2},\nonumber\\
 &=&\frac{\epsilon}{2}+\mathcal O(\epsilon^2).\nonumber
\end{eqnarray}
Both coupling functions are proportional to $F$ and therefore vanish linearly:
\begin{eqnarray}
 \frac{2F(3-R)}{R^3}&=&\mathcal O(\epsilon),\nonumber\\
 \frac{2\Lambda F}{R^2}&=&\mathcal O(\epsilon).\nonumber
\end{eqnarray}
The two amplitudes consequently satisfy the same leading equation,
\begin{eqnarray}
 \frac{\dd^2u_i}{\dd R_*^2}+\varpi^2u_i&=&0,\quad i=2,3. \label{eq.104}
\end{eqnarray}
The elementary solutions are
\begin{eqnarray}
 u_i&\sim&e^{\pm\ii\varpi R_*}.\nonumber
\end{eqnarray}
With the convention $e^{-\ii\omega t}$, the physical future-horizon solution is ingoing,
\begin{eqnarray}
 u_i&\sim&e^{-\ii\varpi R_*}. \label{eq.105}
\end{eqnarray}
Using the Schwarzschild-form relation
\begin{eqnarray}
 R_*&=&R+2\ln\left(\frac{R}{2}-1\right)+{\rm const.},\nonumber
\end{eqnarray}
we find
\begin{eqnarray}
 e^{-\ii\varpi R_*}
 &\propto&\left(\frac{R}{2}-1\right)^{-2\ii\varpi},\nonumber
\end{eqnarray}
so both even amplitudes have the same Frobenius horizon exponent
\begin{eqnarray}
 \boxed{\rho_{\rm h}=-2\ii\varpi}. \label{eq.106}
\end{eqnarray}

\subsubsection{Far-zone massive decay}

For $R\rightarrow\infty$,
\begin{eqnarray}
 F&\rightarrow&1,\nonumber\\
 \frac{2F(3-R)}{R^3}&=&-\frac{2}{R^2}+\mathcal O(R^{-3}),\nonumber\\
 \frac{2\Lambda F}{R^2}&=&\frac{2\Lambda}{R^2}+\mathcal O(R^{-3}).\nonumber
\end{eqnarray}
At leading order both amplitudes satisfy
\begin{eqnarray}
 \frac{\dd^2u_i}{\dd R_*^2}
 +\left(\varpi^2-\mathfrak m^2\right)u_i&=&0. \label{eq.107}
\end{eqnarray}
For a quasibound state we require exponential decay.  Define
\begin{eqnarray}
 q&=&-\sqrt{\mathfrak m^2-\varpi^2},\quad \Re q<0. \label{eq.108}
\end{eqnarray}
Then
\begin{eqnarray}
 u_i&\sim&e^{qR}R^{-\nu}\left[c_i+\mathcal O(R^{-1})\right], \label{eq.109}
\end{eqnarray}
where the Coulomb-type power $\nu$ follows from the next order in the common diagonal equation.  The same calculation as in the axial sector gives
\begin{eqnarray}
 \nu&=&\frac{\varpi^2-q^2}{q}. \label{eq.110}
\end{eqnarray}
The constants $c_2$ and $c_3$ span the two independent even-parity polarizations.  Their subleading ratios distinguish the vector-transverse and longitudinal branches.

\subsubsection{Coordinate-time versus normalized-time threshold}

In the original coordinates the large-radius potential tends to $a\mu^2$, so
\begin{eqnarray}
 \omega_{\rm th}^2&=&a\mu^2.\nonumber
\end{eqnarray}
If the asymptotic time coordinate is normalized by
\begin{eqnarray}
 T&=&\sqrt a\,t,\nonumber
\end{eqnarray}
then the conjugate frequency is
\begin{eqnarray}
 \Omega&=&\frac{\omega}{\sqrt a}.\nonumber
\end{eqnarray}
Consequently
\begin{eqnarray}
 \Omega_{\rm th}&=&\mu. \label{eq.111}
\end{eqnarray}
This distinction is useful when comparing spectral shifts at fixed coordinate parameters with shifts measured by an asymptotically normalized clock.

\begin{figure}[t]
\centering
\includegraphics[width=0.96\columnwidth]{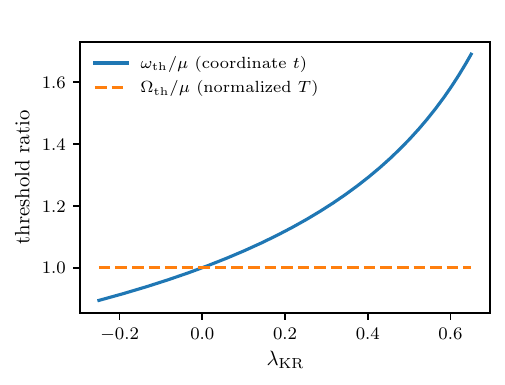}
\caption{Massive threshold in the original coordinate time and in the asymptotically normalized time.  The physical distinction between the two conventions must be kept explicit in any comparison of spectral frequencies. This plot is a normalization comparison, not a claim of a previously unknown mass threshold \cite{RosaDolan2012}.}
\label{fig:thresholdnorm}
\end{figure}

\rev{The matrix Frobenius recurrence is derived in Appendix~\ref{app:matrix}. Its finite-deformation numerical validation remains open; the polar results below use differential-equation matching.}

\subsection{Polarization-resolved criticality: what can and cannot be inferred locally}

The axial mode has a single scalar potential and therefore a sharply defined barrier--well fold.  The even sector does not possess a unique scalar potential before a nonlocal decoupling is performed.  Consequently, one should not define an ``even critical mass'' merely by diagonalizing $\mathbf V_{\rm even}(r)$ at each radius.  Such a local diagonalization depends on the radial field basis and does not preserve the global boundary conditions.

What can be stated exactly is the following.  First, both physical even polarizations share the same massive asymptotic branch point as the axial mode.  Second, the exact map in Eqs.~\eqref{eq.100}--\eqref{eq.103} means that Lorentz violation deforms the even spectral problem through a continuous angular parameter $\Lambda=L/a$ in addition to the rescaled mass and frequency.  Third, the axial fold occurs at
\begin{eqnarray}
 (M\mu)^2_{\rm c,axial}&=&\frac{a^2L}{12},\nonumber\\
 r_{\rm fold}&=&\frac{6M}{a}.\nonumber
\end{eqnarray}
Whether either even branch exhibits a nonanalytic spectral event at the same point is a genuinely spectral question to be decided by Eq.~\eqref{eq.126} or an independent two-channel shooting calculation.

For later numerical work it is useful to define the polarization-resolved spectral functions
\begin{eqnarray}
 \mathcal F_{\rm even}(\omega,\mu,\lKR;\ell)&=&
 \det\left(\boldsymbol\beta_0+\boldsymbol\alpha_0\mathbf R_0\right), \label{eq.127}
\end{eqnarray}
and to follow each simple zero continuously in $(M\mu,\lambda_{\rm KR})$.  A branch crossing or exceptional-point candidate would require simultaneous evidence of root coalescence and loss of simplicity,
\begin{eqnarray}
 \mathcal F_{\rm even}&=&0,\nonumber\\
 \frac{\partial\mathcal F_{\rm even}}{\partial\omega}&=&0. \label{eq.128}
\end{eqnarray}
This criterion is substantially stronger than visual proximity of two complex frequencies.

\subsection{Matrix-Riccati matching}
The two physical even-parity amplitudes satisfy the matrix equation~\eqref{eq.98}. Their exact Kalb--Ramond to Schwarzschild-form map is given by Eq.~\eqref{eq.100}; in particular, the effective centrifugal parameter $\Lambda=L/a$ is continuous. The matrix Frobenius and continued-fraction construction in Appendix~\ref{app:matrix} provides \emph{candidate} complex roots.  A targeted independent validation of the two physically distinct Schwarzschild families and two nearby deformations is presented below; none of the near-degenerate candidates is thereby established as an exceptional point.

An independent numerical representation can be derived directly, without reducing the amplitudes to a scalar potential. Let $\mathbf U$ be a $2\times2$ fundamental matrix whose columns are independent solutions and define the matrix logarithmic derivative $\mathbf Z=\mathbf U'\mathbf U^{-1}$, where a prime here denotes differentiation with respect to the relevant tortoise coordinate. Differentiating gives
\begin{eqnarray}
\mathbf Z'&=&\mathbf U''\mathbf U^{-1}-\mathbf U'\mathbf U^{-1}\mathbf U'\mathbf U^{-1}\nonumber\\
&=&\mathbf V_{\rm even}-\omega^2\mathds{1}_{2}-\mathbf Z^2.\label{eq:matrix-riccati-new}
\end{eqnarray}
For independent ingoing and decaying fundamental matrices, the matching condition on a common nonsingular chart is
\begin{eqnarray}
\det[\mathbf Z_{\rm hor}(r_m;\omega)-\mathbf Z_{\infty}(r_m;\omega)]&=&0.
\label{eq:evans-new}
\end{eqnarray}
These equations are implemented and verified for selected electric-parity roots in Sect.~\ref{sec:verified-polar-riccati}; broad continuation still requires chart-safe propagation. A matrix logarithmic derivative can itself develop poles whenever $\det\mathbf U=0$, so chart switching, QR-based subspace propagation, or a Grassmannian Evans formulation is required for reliable broad sweeps. The corresponding first-order $4\times4$ system supplies the matching subspaces needed for root-multiplicity tests in full detail.

At an exceptional point of the nonlinear spectral condition $D(\omega,\lambda_{\rm KR},M\mu)=0$, necessary candidate conditions are $D=0$ and $\partial_\omega D=0$ on a fixed analytic continuation sheet; eigenvector coalescence and normalization-robust local branching must also be demonstrated. We therefore leave exceptional-point claims explicitly open. Existing exploratory polar tables and graphs are preserved only in the full scientific archive, not interpreted as confirmed results in the present analysis.

\section{Polar quasibound spectra and two-parameter QR--Evans continuation}
\label{sec:consolidated-5}
\subsection{Independent polar quasibound frequencies from matrix Riccati propagation}
\label{sec:verified-polar-riccati}

The two coupled electric-parity amplitudes derived in
Eqs.~\eqref{eq.101}--\eqref{eq.103} now permit a direct numerical test that
does not use the block continued-fraction recursion in the boundary-value
integration.  We integrate the matrix logarithmic derivative
$\mathbf Z=\mathbf U'\mathbf U^{-1}$ on the ingoing and decaying two-dimensional
solution subspaces. A prime in this section means $d/dR_*$ and
$F=1-2/R$. The implemented first-order equation in the area coordinate is
\begin{eqnarray}
 \frac{d\mathbf Z}{dR}&=&\frac{\mathbf V_{\rm even}(R)
 -\varpi^2\mathds{1}_{2}-\mathbf Z^2}{F(R)}.
 \label{eq:polar-riccati-implemented}
\end{eqnarray}
The boundary matrices are constructed independently from (i) the ingoing
horizon Frobenius series, with exponent $\rho=-2i\varpi$, and (ii) the
decaying large-$R$ asymptotic series
$e^{-\kappa R}R^p\sum_{n\ge0}\mathbf B_nR^{-n}$, where
\begin{eqnarray}
 \kappa&=&\sqrt{\mathfrak m^2-\varpi^2},\quad
 \operatorname{Re}\kappa>0,\quad
 p=\frac{\mathfrak m^2}{\kappa}-2\kappa.
 \label{eq:polar-asymptotic-implemented}
\end{eqnarray}
The two arbitrary leading amplitude vectors are assembled into a
$2\times2$ fundamental matrix. Matching at $R_m$ then requires
\begin{eqnarray}
 D_{\rm R}(\varpi)&=&\det[\mathbf Z_{\rm h}(R_m)
                 -\mathbf Z_\infty(R_m)]=0.
 \label{eq:polar-riccati-residual}
\end{eqnarray}
The horizon and infinity boundary data are constructed from truncated
Frobenius and asymptotic expansions, respectively, with coefficients fixed
by substitution into the coupled radial equations. The initial guess for
the weakly damped branch is informed by the Schwarzschild block-recurrence
benchmark. The reported root residuals, radial integration, and
boundary-subspace matching are evaluated independently of that recurrence.

\begin{table*}[t]
\centering
\caption{Independently integrated polar quasibound frequencies for
$\ell=1$, $M\mu=0.4$. The two Schwarzschild families $\mathcal E_-$ and
$\mathcal E_+$ are distinguished by their established benchmark and very
different damping rates. The first nonzero-deformation calculations are
continued from these branches. Numerical values are diagnostics at the stated
tolerances, not certified digits. The last column gives the smallest singular
value of an independent $4\times4$ QR-propagated matching matrix.}
\label{tab:polar-riccati-verified}
\begin{tabular}{cccccc}
\toprule
$\lambda_{\rm KR}$ & branch & $\operatorname{Re}(M\omega)$ & $-\operatorname{Im}(M\omega)$ & $\operatorname{Re}(M\Omega)$ & $\sigma_{\min}^{\rm QR}$\\
\midrule
$0$ & $\mathcal E_-$ & $0.360779151$ & $2.1668628\times10^{-2}$ & $0.360779151$ & $1.75\times10^{-12}$\\
$0$ & $\mathcal E_+$ & $0.397775602$ & $1.5810732\times10^{-6}$ & $0.397775602$ & $1.26\times10^{-12}$\\
$0.02$ & $\mathcal E_-$ & $0.365319852$ & $2.0611225\times10^{-2}$ & $0.361648202$ & $1.04\times10^{-12}$\\
$0.02$ & $\mathcal E_+$ & $0.401945688$ & $1.0889833\times10^{-6}$ & $0.397905930$ & $1.35\times10^{-12}$\\
$0.04$ & $\mathcal E_-$ & $0.370020765$ & $1.9559837\times10^{-2}$ & $0.362544828$ & $7.41\times10^{-13}$\\
$0.04$ & $\mathcal E_+$ & $0.406237839$ & $7.5032868\times10^{-7}$ & $0.398030168$ & $1.61\times10^{-12}$\\
\bottomrule
\end{tabular}
\end{table*}

A second numerical representation propagates the complete four-component
first-order system without integrating the nonlinear Riccati matrix.
Each $4\times2$ solution basis is periodically QR-orthonormalized, and the
smallest singular value of the joined $4\times4$ horizon/infinity subspaces
is evaluated at the independently determined Riccati roots. This
\emph{separate numerical propagation} confirms the local matching in
Table~\ref{tab:polar-riccati-verified}, although it shares the same field
equations and asymptotic boundary expansions. The Schwarzschild lower
branch at $M\mu=0.2$ also reproduces the original block-continued-fraction
benchmark, $M\omega\simeq0.195325643307022-1.06451361289097\times10^{-4}i$,
to an absolute difference of approximately $6.4\times10^{-15}$ in one
high-tolerance run. This last difference should \emph{not} be interpreted
as a rigorous fifteen-digit error bound.

The sensitivity analysis comprises nine boundary/series configurations for each
of six roots (54 tests). The maximum absolute determinant evaluated at the
reference roots is $5.4\times10^{-12}$, while the independent QR subspace
singular values in Table~\ref{tab:polar-riccati-verified} remain of order
$10^{-12}$. All six roots obey the selected decaying-sheet condition
$\operatorname{Re}\kappa>0$. The direct matrix-Riccati chart may become
singular away from this parameter range, however, when
$\det\mathbf U=0$. These six reference poles anchor the two-parameter continuation below.

\subsection{QR--Evans matching and local multiplicity diagnostics}\label{sec:polar-qr-v20}\label{sec:epbounded}\label{sec:evans-holo-v21}
\rev{The purpose of the polar survey is to test whether the nearby axial geometric fold induces a visible anomaly in the coupled electric spectrum, rather than to presume a collision of two widely separated poles. The labels $\mathcal E_-$ and $\mathcal E_+$ distinguish strongly and weakly damped branches connected to Schwarzschild electric benchmarks. The usual $S=\pm1$ classification of Rosa--Dolan describes weak-coupling families; an explicit continuation to that limit is needed to establish the assignment and longitudinal/transverse content of each branch here.}

For the two-component polar amplitude $\bm u=(u_2,u_3)^\mathsf T$, introduce $\bm p=\partial_{R_*}\bm u$. In the exact Schwarzschild-form coordinate $R$, with $F=1-2/R$, the coupled wave system is
\begin{eqnarray}
\frac{\dd}{\dd R}\begin{pmatrix}\bm u\\\bm p\end{pmatrix}
&=&\frac1F\begin{pmatrix}
0&\mathds{1}_{2}\\ \bm V_{\rm pol}-\varpi^2\mathds{1}_{2}&0
\end{pmatrix}\begin{pmatrix}\bm u\\\bm p\end{pmatrix}.\label{eq:v20-firstorder}
\end{eqnarray}
where $\mathds{1}_{2}=\mathrm{diag}(1,1)$ is the two-dimensional identity and $\bm V_{\rm pol}$ is the matrix already defined in the polar reduction. Orthonormal bases $Q_{\rm H},Q_{\infty}\in\mathbb C^{4\times2}$ span the ingoing-horizon and decaying-infinity solution spaces. Repeated QR orthogonalizations maintain stable bases during the integrations, without requiring the amplitude block to remain invertible. The matching condition becomes
\begin{eqnarray}
\mathcal E_{\rm QR}(\omega)&=&\det[Q_{\rm H}\;Q_{\infty}]=0,\quad
\sigma_{\min}([Q_{\rm H}\;Q_{\infty}])=0.
\label{eq:v20-evans}
\end{eqnarray}
Under nonsingular changes of boundary basis, zeros of this determinant are unchanged. Because QR uses conjugate transpose, the \emph{normalized} determinant is not generally holomorphic in the frequency; its complex derivative must therefore not be used as a proof of multiple analytic Evans zeros or exceptional points. Near a simple root its vanishing, together with independent complex-frequency and subspace checks, nevertheless provides a useful computational condition.

Let $\bm u=(u_2,u_3)^T$ and let $R_*$ denote the Schwarzschild-form tortoise coordinate. The coupled polar system is
\begin{eqnarray}
\bm u''&=&\bigl[\bm V_{\rm even}(R)-\varpi^2\mathds{1}_{2}\bigr]\bm u,
\quad
\varpi=\frac{M\omega}{a^2},
\end{eqnarray}
where primes are derivatives with respect to $R_*$. For a nonsingular $2\times2$ fundamental matrix $\bm U$ define $\bm Z=\bm U'\bm U^{-1}$. Its Riccati equation is
\begin{eqnarray}
\label{eq:v21-riccati}
\frac{d\bm Z}{dR_*}&=&\bm V_{\rm even}-\varpi^2\mathds{1}_{2}-\bm Z^2.
\end{eqnarray}
The two boundary subspaces produce $\bm Z_{\rm H}$ and $\bm Z_\infty$ at a common point $R_m$. Within the common regular chart, set
\begin{eqnarray}
 D_{\rm loc}(W)&=&\det[\bm Z_{\rm H}(W;R_m)
 -\bm Z_\infty(W;R_m)],\nonumber\\
 W&=&M\omega\equiv w.\label{eq:v21-evans}
\end{eqnarray}
The Frobenius boundary matrices depend analytically on $W$ away from recurrence singularities and the mass-shell branch point. The nonlinear ODE preserves local holomorphic dependence until a Riccati-chart singularity occurs. Thus $D_{\rm loc}$ is locally holomorphic under these explicit assumptions.

Set $\bm S=\partial_W\bm Z$. Differentiating Eq.~\eqref{eq:v21-riccati} \emph{at fixed background parameters} and using $\partial_W\varpi=a^{-2}$ gives the linear matrix variational equation
\begin{eqnarray}
\label{eq:v21-variation}
\frac{d\bm S}{dR_*}&=&-\frac{2\varpi}{a^2}\mathds{1}_{2}-\bm S\bm Z-\bm Z\bm S.
\end{eqnarray}
We propagate $\bm Z$ and $\bm S$ simultaneously, with the boundary derivatives initialized by symmetric complex-frequency differences of the \emph{analytic Frobenius matrices}, not by differentiating the final Evans determinant. With $\bm\Delta=\bm Z_{\rm H}-\bm Z_\infty$ and $\bm\Delta_W=\bm S_{\rm H}-\bm S_\infty$, the derivative is evaluated explicitly as
\begin{eqnarray}
D_{\rm loc}'&=&(\Delta_W)_{11}\Delta_{22}+\Delta_{11}(\Delta_W)_{22}\nonumber\\
&&-(\Delta_W)_{12}\Delta_{21}-\Delta_{12}(\Delta_W)_{21}.\label{eq:v21-detder}
\end{eqnarray}
No division by $D_{\rm loc}$ is required at a zero. A necessary condition for a double zero is $D_{\rm loc}=D_{\rm loc}'=0$; its interpretation as a physical exceptional point would additionally require a one-dimensional eigenspace and a generalized eigenvector or a local square-root unfolding.

\subsection{Both electric families on the two-parameter grid}\label{sec:v22-grid}\label{sec:v23-both}
\rev{At $\ell=1$ we follow both branches on the grid $M\mu\in\{0.38,0.39,0.40,0.41,0.42\}$ and $\lambda_{\rm KR}\in\{0,0.025,0.05,0.075,0.10\}$. Predictors are corrected by QR matching, enforcing the decaying sheet, frequency continuity and overlap of consecutive reconstructed matching states. The weakly damped path uses steps of $0.005$ in both parameters from the reference seed $(0.40,0.05)$. The 25 lower-branch Riccati matches have residuals at most $1.76\times10^{-10}$; nine QR checks have smallest singular values at most $1.25\times10^{-10}$. For the upper branch, the maximum QR singular value is $5.03\times10^{-10}$ and the minimum consecutive-state overlap is $0.9999859$. Configuration changes shift the selected upper-branch frequencies by at most $6.47\times10^{-14}$ and the tested lower-branch frequencies by at most $5.21\times10^{-12}$.}

\begin{figure*}[t]
\centering\includegraphics[width=.78\textwidth]{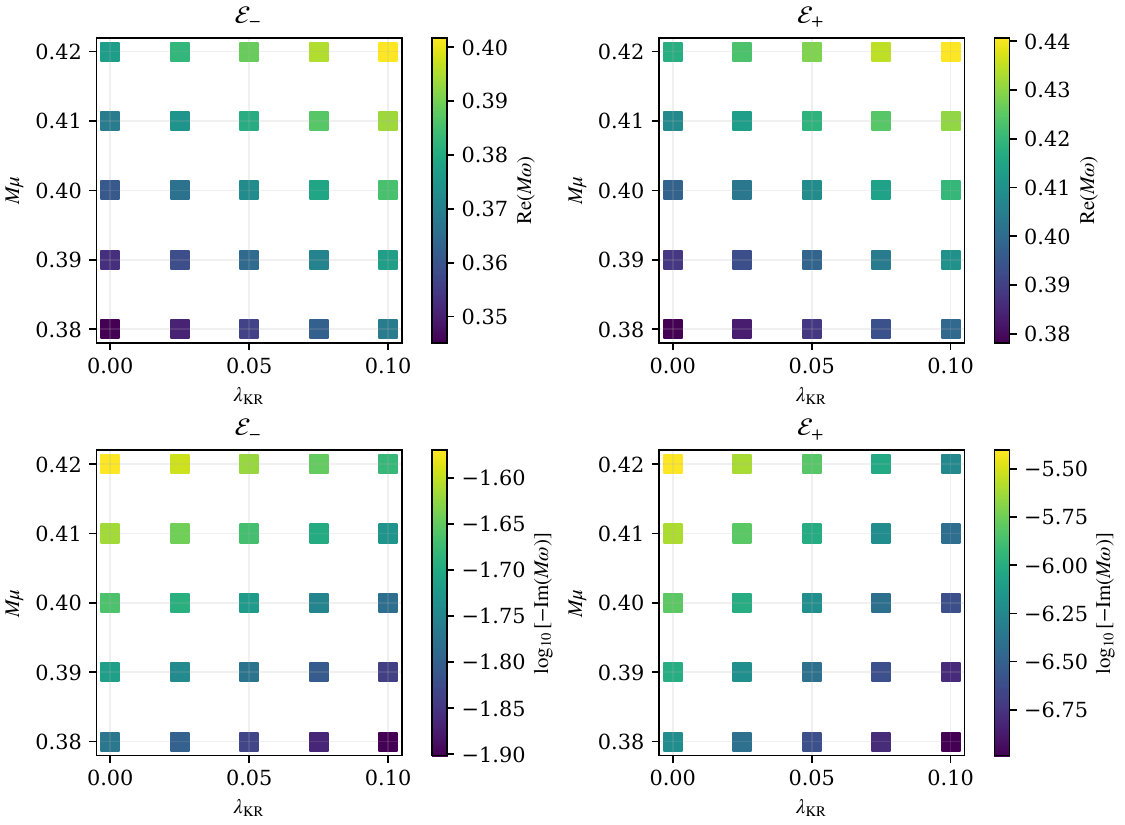}
\caption{\rev{Both electric families at the same 25 directly calculated nodes. Colors show coordinate frequency and logarithmic damping; no interpolated eigenvalues are implied.}}
\label{fig:raul-polar-grid}
\end{figure*}
\rev{The minimum sampled separation is $\min|M\omega_+-M\omega_-|=0.03305048$. The local holomorphic derivatives are nonzero at the nine audited nodes of each family. At the lower-branch nodes the smallest observed $|\partial_W D_{\rm loc}|$ is $4.21$ in the chosen chart. The upper branch also has nonzero derivatives along the nineteen-point deformation path at $M\mu=0.4$. The axial fold crosses this mass window only for $0\leq\lambda_{\rm KR}\lesssim0.0280$; the remaining nodes sample its vicinity on the subcritical side. No polar coalescence is resolved on the sampled grid.}

\subsection{Scope of the numerical evidence}
\rev{Finite-grid minima, matching residuals, truncation comparisons and configuration sensitivities are numerical diagnostics rather than certified bounds. Riccati and QR propagation share the boundary expansions; the block continued fraction has not been independently validated for finite-deformation polar modes. Local holomorphic tests require regular charts and a fixed decay sheet. Neither the selected axial associations nor the polar grid provide a complete spectrum or a continuous-domain exclusion of exceptional points.}

\rev{The present quasibound calculations impose decay at infinity, unlike outgoing quasinormal resonances or real-frequency transmission. Related WKB and scattering studies~\cite{KonoplyaWKB2019,GuoKRgreybody2024,BaruahKRgreybody2025,BertiCardosoStarinets2009} address these distinct boundary conditions; no thermal or outgoing spectrum is needed for the conclusions here.}

\section{Conclusions}
\label{sec:consolidated-7}
\label{sec:conclusions}

We have studied Maxwell--Proca test fields on the static Lorentz-violating Kalb--Ramond geometry, focusing on the decaying quasibound spectra of both parity sectors. The background metric and the Schwarzschild Proca formulation are established inputs. Our model-specific analytical results are the exact deformation-dependent radial reduction and the correspondence between the normalized axial massive-vector potential and the timelike-geodesic effective potential.

The axial barrier maximum and outer well minimum coalesce at
\begin{eqnarray}
 r_{\rm fold}&=&6M(1-\lambda_{\rm KR}),\nonumber\\
 (M\mu_{\rm c})^2&=&\frac{\ell(\ell+1)}{12(1-\lambda_{\rm KR})^2}.
\end{eqnarray}
The fold radius coincides with the known marginally stable circular orbit in this geometry. The local fold exponents are generic; their explicit coefficients and their relationship to the massive-vector spectral parameters are the relevant model-dependent information. In particular, coalescence of extrema of a \emph{real} potential is not equivalent to a multiple root of the complex-frequency spectral condition.

\rev{For eight selected axial critical cases, the real-energy gaps are positive and the barrier crossings occur at $0.8929\leq m_\times/m_c\leq0.9427$. The ratios decrease with deformation for the sampled multipoles, while normalized-time damping increases between crossing and fold. Bilateral derivatives at two truncations support local spectral regularity. Independent Riccati shooting recovers the poles at all eight crossing couplings in two configurations, with a maximum frequency difference of $5.46\times10^{-14}$. The weak-binding associations of the selected axial branches are $n=0,1,0$ for $\ell=1,2,3$; no complete enumeration is implied. The angular parameter $\Lambda=L/a$ distinguishes deformation-dependent changes in the mapped equations from frequency and radial rescalings.}

\rev{In the coupled polar problem, both electric families are recovered on the 25-node parameter grid. The minimum sampled gap $|M\omega_+-M\omega_-|=0.03305048$, together with local Evans checks, gives no confirmed exceptional point; it neither bounds the continuous rectangle nor excludes collisions with untracked roots. At fixed lapse mass and normalized time, the angular parameter isolates the residual deformation in both parity maps. These results establish parameter-bounded spectral comparisons rather than a universal criticality or completeness theorem.}

\section*{Data availability}
\rev{Data supporting the figures and tables are available from the corresponding author upon reasonable request.}

\section*{Code availability}
\rev{Numerical implementations and reproduction instructions are available from the corresponding author upon reasonable request.}

\section*{Competing interests}
The authors declare no competing interests.

\section*{Use of AI tools}
ChatGPT (OpenAI) was used as an assistive tool for English-language editing, manuscript organization, LaTeX formatting, literature searches and bibliographic checks, drafting intermediate mathematical derivations, computational prototyping, and development and review of numerical workflows. AI-assisted outputs were subject to review by the authors; symbolic relations and selected numerical results were rechecked separately as described in the text. The authors retain full responsibility for the validity, originality, reproducibility, citations, and conclusions of the submitted work. No AI system is listed as an author.

\clearpage
\appendix
\section{Matrix Frobenius recurrence}\label{app:matrix}

\label{sec:matrixfrobenius}

The exact Schwarzschild-form reduction makes it possible to construct a matrix-valued Frobenius problem with precisely the same analytic endpoints as in Schwarzschild.  We give the construction in a form suitable for symbolic or numerical implementation.

\subsection{Compact coordinate and common asymptotic factor}

Introduce
\begin{eqnarray}
 z&=&1-\frac{2}{R}. \label{eq.112}
\end{eqnarray}
Then
\begin{eqnarray}
 R&=&\frac{2}{1-z},\nonumber\\
 F&=&z,\nonumber\\
 \frac{\dd R}{\dd z}&=&\frac{2}{(1-z)^2}.\nonumber
\end{eqnarray}
Since
\begin{eqnarray}
 \frac{\dd}{\dd R_*}&=&F\frac{\dd}{\dd R},\nonumber
\end{eqnarray}
we obtain
\begin{eqnarray}
 \frac{\dd}{\dd R_*}
 &=&\frac{z(1-z)^2}{2}\frac{\dd}{\dd z}.\nonumber
\end{eqnarray}
Define
\begin{eqnarray}
 h(z)&=&\frac{z(1-z)^2}{2}.\nonumber
\end{eqnarray}
Then
\begin{eqnarray}
 \frac{\dd^2}{\dd R_*^2}
 &=&h^2\frac{\dd^2}{\dd z^2}+hh'\frac{\dd}{\dd z}. \label{eq.113}
\end{eqnarray}
The common diagonal potential becomes
\begin{eqnarray}
 V_0(z)
 &=&z\left[\mathfrak m^2+\frac{\Lambda}{4}(1-z)^2\right].\nonumber
\end{eqnarray}
The two couplings are polynomial functions times the same endpoint factors:
\begin{eqnarray}
 C(z)
 &=&\frac{z(1-3z)(1-z)^2}{4},\nonumber\\
 D(z)
 &=&\frac{\Lambda z(1-z)^2}{2}. \label{eq.114}
\end{eqnarray}
Thus the coupled equations are
\begin{eqnarray}
 h^2u_{2,zz}+hh'u_{2,z}+[\varpi^2-V_0+C]u_2-Cu_3&=&0,\nonumber\\
 h^2u_{3,zz}+hh'u_{3,z}+[\varpi^2-V_0]u_3+Du_2&=&0. \label{eq.115}
\end{eqnarray}

Factor out the exact horizon and infinity behavior,
\begin{eqnarray}
 \mathbf u(z)
 &=&P(z)\sum_{n=0}^{\infty}\mathbf a_nz^n,\nonumber\\
 \mathbf u&=&\left(\begin{array}{c}u_2\\u_3\end{array}\right),\quad
 \mathbf a_n=\left(\begin{array}{c}a_n^{(2)}\\a_n^{(3)}\end{array}\right),\nonumber\\
 P(z)&=&z^{\rho}(1-z)^{\nu}\exp\left(\frac{2q}{1-z}\right),\nonumber\\
 \rho&=&-2\ii\varpi. \label{eq.116}
\end{eqnarray}
The logarithmic derivatives of $P$ are
\begin{eqnarray}
 \frac{P'}{P}
 &=&\frac{\rho}{z}-\frac{\nu}{1-z}+\frac{2q}{(1-z)^2},\nonumber\\
 \frac{P''}{P}
 &=&\left(\frac{P'}{P}\right)^2
 -\frac{\rho}{z^2}-\frac{\nu}{(1-z)^2}+\frac{4q}{(1-z)^3}. \label{eq.117}
\end{eqnarray}
Writing
\begin{eqnarray}
 \mathbf u&=&P\mathbf y,\nonumber
\end{eqnarray}
we have
\begin{eqnarray}
 \mathbf u_z&=&P\left(\mathbf y_z+\frac{P'}{P}\mathbf y\right),\nonumber\\
 \mathbf u_{zz}&=&P\left[\mathbf y_{zz}+2\frac{P'}{P}\mathbf y_z+\frac{P''}{P}\mathbf y\right].\nonumber
\end{eqnarray}
After division by $P$, Eq.~\eqref{eq.115} takes the compact matrix form
\begin{eqnarray}
 \mathbf P_2(z)\mathbf y_{zz}
 +\mathbf P_1(z)\mathbf y_z
 +\mathbf P_0(z)\mathbf y&=&0, \label{eq.118}
\end{eqnarray}
where
\begin{eqnarray}
 \mathbf P_2(z)&=&h^2\mathds{1}_{2},\nonumber\\
 \mathbf P_1(z)&=&\left[2h^2\frac{P'}{P}+hh'\right]\mathds{1}_{2},\nonumber\\
 \mathbf P_0(z)&=&\left[h^2\frac{P''}{P}+hh'\frac{P'}{P}+\varpi^2-V_0\right]\mathds{1}_{2}\nonumber\\
 &&+\left(\begin{array}{cc}C&-C\\D&0\end{array}\right). \label{eq.119}
\end{eqnarray}
The choices of $\rho$, $q$, and $\nu$ cancel the singular terms associated with the horizon and the massive infinity.  Multiplying Eq.~\eqref{eq.118} by the minimal common denominator $z^2(1-z)^4$ leaves a finite polynomial matrix differential equation.  We denote the resulting polynomial coefficients by
\begin{eqnarray}
 \widetilde{\mathbf P}_k(z)
 &=&\sum_{j=0}^{J_k}\mathbf P_{k,j}z^j,\quad k=0,1,2. \label{eq.120}
\end{eqnarray}
Substitution of
\begin{eqnarray}
 \mathbf y&=&\sum_{n=0}^{\infty}\mathbf a_nz^n,\nonumber\\
 \mathbf y_z&=&\sum_{n=1}^{\infty}n\mathbf a_nz^{n-1},\nonumber\\
 \mathbf y_{zz}&=&\sum_{n=2}^{\infty}n(n-1)\mathbf a_nz^{n-2},\nonumber
\end{eqnarray}
then gives, coefficient by coefficient,
{\begin{eqnarray}
&&\sum_{j=0}^{J_2}(n-j+2)(n-j+1)\mathbf P_{2,j}\mathbf a_{n-j+2}\nonumber\\
&&+\sum_{j=0}^{J_1}(n-j+1)\mathbf P_{1,j}\mathbf a_{n-j+1}\nonumber\\
&&+\sum_{j=0}^{J_0}\mathbf P_{0,j}\mathbf a_{n-j}=0.\label{eq.121}
\end{eqnarray}}
Equation~\eqref{eq.121} is the explicit block-Frobenius recurrence before Gaussian reduction.  Terms with negative subscripts are omitted.  Because the coefficient matrices have finite polynomial degree, the recurrence has finite bandwidth.

\subsection{Reduction to a three-term block recurrence}

Suppose the raw recurrence generated by Eq.~\eqref{eq.121} has the form
\begin{eqnarray}
 \sum_{j=0}^{p}\mathbf C_{n,j}\mathbf a_{n+1-j}&=&0. \label{eq.122}
\end{eqnarray}
A block Gaussian elimination can remove the most distant coefficient successively.  For a four-term stage,
\begin{eqnarray}
 \mathbf A_n\mathbf a_{n+1}
 +\mathbf B_n\mathbf a_n
 +\mathbf C_n\mathbf a_{n-1}
 +\mathbf D_n\mathbf a_{n-2}&=&0,\nonumber
\end{eqnarray}
one uses the previous reduced equation to eliminate $\mathbf a_{n-2}$.  If
\begin{eqnarray}
 \widetilde{\mathbf A}_{n-1}\mathbf a_n
 +\widetilde{\mathbf B}_{n-1}\mathbf a_{n-1}
 +\widetilde{\mathbf C}_{n-1}\mathbf a_{n-2}&=&0,\nonumber
\end{eqnarray}
then
\begin{eqnarray}
 \mathbf a_{n-2}
 &=&-\widetilde{\mathbf C}_{n-1}^{-1}
 \left(\widetilde{\mathbf A}_{n-1}\mathbf a_n
 +\widetilde{\mathbf B}_{n-1}\mathbf a_{n-1}\right).\nonumber
\end{eqnarray}
Substitution gives the reduced coefficients
\begin{eqnarray}
 \widetilde{\mathbf A}_n&=&\mathbf A_n,\nonumber\\
 \widetilde{\mathbf B}_n&=&\mathbf B_n-\mathbf D_n\widetilde{\mathbf C}_{n-1}^{-1}\widetilde{\mathbf A}_{n-1},\nonumber\\
 \widetilde{\mathbf C}_n&=&\mathbf C_n-\mathbf D_n\widetilde{\mathbf C}_{n-1}^{-1}\widetilde{\mathbf B}_{n-1}. \label{eq.123}
\end{eqnarray}
Repeated application reduces any finite-band recurrence to the matrix three-term form
\begin{eqnarray}
 \boxed{\boldsymbol\alpha_n\mathbf a_{n+1}
 +\boldsymbol\beta_n\mathbf a_n
 +\boldsymbol\gamma_n\mathbf a_{n-1}=0}. \label{eq.124}
\end{eqnarray}

\subsection{Matrix continued fraction}

Define a transfer matrix $\mathbf R_n$ by
\begin{eqnarray}
 \mathbf a_{n+1}&=&\mathbf R_n\mathbf a_n.\nonumber
\end{eqnarray}
Using Eq.~\eqref{eq.124} at index $n+1$,
\begin{eqnarray}
 \boldsymbol\alpha_{n+1}\mathbf a_{n+2}
 +\boldsymbol\beta_{n+1}\mathbf a_{n+1}
 +\boldsymbol\gamma_{n+1}\mathbf a_n&=&0.\nonumber
\end{eqnarray}
With
\begin{eqnarray}
 \mathbf a_{n+2}&=&\mathbf R_{n+1}\mathbf a_{n+1},\nonumber
\end{eqnarray}
we obtain
\begin{eqnarray}
 \left(\boldsymbol\alpha_{n+1}\mathbf R_{n+1}+\boldsymbol\beta_{n+1}\right)\mathbf a_{n+1}
 & =&-\boldsymbol\gamma_{n+1}\mathbf a_n,\nonumber
\end{eqnarray}
and therefore
\begin{eqnarray}
 \boxed{\mathbf R_n
 =-\left(\boldsymbol\beta_{n+1}+\boldsymbol\alpha_{n+1}\mathbf R_{n+1}\right)^{-1}\boldsymbol\gamma_{n+1}}. \label{eq.125}
\end{eqnarray}
The $n=0$ equation supplies the spectral condition.  If $\boldsymbol\gamma_0=0$,
\begin{eqnarray}
 \left(\boldsymbol\beta_0+\boldsymbol\alpha_0\mathbf R_0\right)\mathbf a_0&=&0.\nonumber
\end{eqnarray}
A nontrivial vector $\mathbf a_0$ exists only if
\begin{eqnarray}
 \boxed{\det\left(\boldsymbol\beta_0+\boldsymbol\alpha_0\mathbf R_0\right)=0}. \label{eq.126}
\end{eqnarray}
Equation~\eqref{eq.126} is the matrix-valued analogue of Leaver's scalar continued-fraction equation.  The numerical implementation can be stabilized by backward recursion from a large truncation order and by monitoring the smallest singular value of the matrix in Eq.~\eqref{eq.126} rather than its determinant alone.

The important conceptual point is that all dependence on the Kalb--Ramond deformation enters the reduced spectral problem through the effective parameters $(\varpi,\mathfrak m,\Lambda)$.  This provides a direct route for adapting and benchmarking Schwarzschild matrix-continuued-fraction codes.


\clearpage
\begin{thebibliography}{99}
\bibitem{Konoplya2006}
R. A. Konoplya,
``Massive vector field perturbations in the Schwarzschild background: Stability and quasinormal spectrum,''
Phys. Rev. D \textbf{73}, 024009 (2006), DOI: 10.1103/PhysRevD.73.024009.
\bibitem{RosaDolan2012}
J. G. Rosa and S. R. Dolan,
``Massive vector fields on the Schwarzschild spacetime: Quasinormal modes and bound states,''
Phys. Rev. D \textbf{85}, 044043 (2012), DOI: 10.1103/PhysRevD.85.044043; arXiv:1110.4494 [hep-th].
\bibitem{Ge2025}
X.-H. Ge, M. Matsumoto, and K. Zhang,
``Massive vector field perturbations in the Schwarzschild spacetime from supersymmetric gauge theory,''
Phys. Rev. D \textbf{112}, 046024 (2025), DOI: 10.1103/3dwl-gg8g; arXiv:2502.15627 [hep-th].
\bibitem{Yang2023}
K. Yang, Y.-Z. Chen, Z.-Q. Duan, and J.-Y. Zhao,
``Static and spherically symmetric black holes in gravity with a background Kalb--Ramond field,''
Phys. Rev. D \textbf{108}, 124004 (2023), DOI: 10.1103/PhysRevD.108.124004; arXiv:2308.06613 [gr-qc].
\bibitem{Junior2024}
E. L. B. Junior, J. T. S. S. Junior, F. S. N. Lobo, M. E. Rodrigues, D. Rubiera-Garcia, L. F. Dias da Silva, and H. A. Vieira,
``Spontaneous Lorentz symmetry-breaking constraints in Kalb--Ramond gravity,''
Eur. Phys. J. C \textbf{84}, 1257 (2024), DOI: 10.1140/epjc/s10052-024-13619-3; arXiv:2405.03291 [gr-qc].
\bibitem{Periodic2025}
E. L. B. Junior et al.,
``Periodical orbits and waveforms with spontaneous Lorentz symmetry-breaking in Kalb--Ramond gravity,''
Eur. Phys. J. C \textbf{85}, 557 (2025), DOI: 10.1140/epjc/s10052-025-14299-3.
\bibitem{Deng2025}
W. Deng, W. Liu, K. Xiao, and J. Jing,
``Quasinormal modes of scalar, electromagnetic, and gravitational perturbations in slowly rotating Kalb--Ramond black holes,''
Eur. Phys. J. C \textbf{86}, 232 (2026), DOI: 10.1140/epjc/s10052-026-15470-0; arXiv:2511.19553 [gr-qc].
\bibitem{GuKR2025}
Y.-T. Gu, W.-D. Guo, and Y.-X. Liu,
``Quasinormal modes of an electrically charged Kalb--Ramond black hole,''
Chin. Phys. C \textbf{50}, 105102 (2026), DOI: 10.1088/1674-1137/ae6ed3; arXiv:2509.23732 [gr-qc].
\bibitem{AraujoDyonic2026}
A. A. Ara\'ujo Filho,
``Perturbative dynamics and relativistic effects of a dyonic Kalb--Ramond black hole,''
arXiv:2605.28580 [gr-qc] (2026).
\bibitem{Murodov2026}
S. Murodov, O. Kholturayev, Bahodir Ahmedov, B. Rahmatov, I. Egamberdiev, and Bobomurat Ahmedov,
``Massive neutral Dirac quasibound states in a Newman--Janis-generated rotating charged Kalb--Ramond black-hole geometry,''
arXiv:2608.09313 [gr-qc] (2026).
\bibitem{AydemirSakalli2026}
G. Aydemir and I. Sakall\i,
``Stationary scalar clouds, quasinormal ringing, and observational signatures of (near-)extremal rotating Kalb--Ramond black holes: from the superradiant threshold to EHT bounds,''
Eur. Phys. J. C \textbf{86}, 1222 (2026), doi:10.1140/epjc/s10052-026-16352-1.
\bibitem{Hod2025Inverted}
S. Hod, ``Bound-state resonances of the Schwarzschild black hole: Analytic treatment,'' Phys. Rev. D \textbf{112}, 064040 (2025), DOI: 10.1103/l5dw-ddrw; arXiv:2506.15768.
\bibitem{Hod2026Scalar}
\rev{S. Hod, ``Long-lived resonances of massive scalar fields in the Reissner--Nordstr\"om black-hole spacetime: Analytic treatment in the large-mass regime,'' Phys. Lett. B \textbf{874}, 140228 (2026), arXiv:2511.10521.}
\bibitem{RibeiroSchwarzschild}
\rev{A. S. Ribeiro and F. A. P. Alves Junior, ``Geometric criticality and regular quasibound spectra of axial Proca fields in Schwarzschild spacetime,'' arXiv:2610.34528 (2026).}
\bibitem{LiVlasov2024}
X.-X. Zhang and Y.-H. Jiang,
``Accretion of Vlasov gas onto a black hole in the Kalb--Ramond field,''
Chin. Phys. C \textbf{49}, 025101 (2025), DOI: 10.1088/1674-1137/ad8bb0.
\bibitem{ChargedKR2024}
Z.-Q. Duan, J.-Y. Zhao, and K. Yang,
``Electrically charged black holes in gravity with a background Kalb--Ramond field,''
Eur. Phys. J. C \textbf{84}, 798 (2024), DOI: 10.1140/epjc/s10052-024-13188-5; arXiv:2310.13555 [gr-qc].
\bibitem{KonoplyaWKB2019} R. A. Konoplya, A. Zhidenko, and A. F. Zinhailo, ``Higher order WKB formula for quasinormal modes and grey-body factors: recipes for quick and accurate calculations,'' Class. Quantum Grav. \textbf{36}, 155002 (2019), DOI: 10.1088/1361-6382/ab2e25; arXiv:1904.10333.
\bibitem{GuoKRgreybody2024} W.-D. Guo et al., ``Quasinormal modes and greybody factor of a Lorentz-violating black hole,'' JCAP \textbf{07} (2024) 008, doi:10.1088/1475-7516/2024/07/008.
\bibitem{BaruahKRgreybody2025} A. Baruah et al., ``Quasinormal modes, greybody factors, and Hawking radiation sparsity of black holes influenced by a global monopole charge in Kalb--Ramond gravity,'' JCAP \textbf{08} (2025) 023, doi:10.1088/1475-7516/2025/08/023.
\bibitem{BertiCardosoStarinets2009} E. Berti, V. Cardoso, and A. O. Starinets, ``Quasinormal modes of black holes and black branes,'' Class. Quantum Grav. \textbf{26}, 163001 (2009), doi:10.1088/0264-9381/26/16/163001; arXiv:0905.2975 [gr-qc].
\end{thebibliography}
\end{document}